\documentclass[twocolumn]{aastex63}

\usepackage{apjfonts}
\usepackage{lineno}
\usepackage{color}
\usepackage{soul}
\usepackage{natbib}
\usepackage{hyperref}
\usepackage{amsmath}
\usepackage{xspace}
\usepackage{graphicx}
\usepackage{enumitem}
\usepackage{amssymb}
\usepackage{xifthen}
\usepackage[normalem]{ulem}
\usepackage{aas_macros}
\usepackage{amsfonts}
\usepackage{amsmath}
\usepackage{multirow}
\usepackage{dsfont}
\usepackage{hyperref}
\usepackage{lineno}

\newcommand{\code}[1]{\texttt{#1}\xspace}

\hypersetup{    
  colorlinks      = {true},
  linkcolor       = {blue},
  citecolor       = {blue},
  urlcolor        = {blue},
}

\newcommand{\tucv}{Tuc~V}

\shorttitle{Tucana V}
\shortauthors{}

\begin{document}
\title{Chemical Diversity on Small Scales - Abundance Analysis of the Tucana V Ultra-Faint Dwarf Galaxy}

\author[0000-0001-6154-8983]{Terese T.~Hansen}
\affil{Department of Astronomy, Stockholm University, AlbaNova
University Center, SE-106 91 Stockholm, Sweden}

\author[0000-0002-4733-4994]{Joshua D.~Simon}
\affil{Observatories of the Carnegie Institution for Science, 813 Santa Barbara St., Pasadena, CA 91101, USA}

\author[0000-0002-9110-6163]{Ting~S.~Li}
\affiliation{Department of Astronomy and Astrophysics, University of Toronto, 50 St. George Street, Toronto ON, M5S 3H4, Canada}

\author{Domani Sharkey}
\affil{Department of Astronomy, California Institute of Technology, 1200 E. California Blvd., Pasadena, CA 91125, USA}

\author[0000-0002-4863-8842]{Alexander~P.~Ji}
\affiliation{Department of Astronomy \& Astrophysics, University of Chicago, 5640 S Ellis Avenue, Chicago, IL 60637, USA}

\author{Ian B. Thompson}
\affil{Observatories of the Carnegie Institution for Science, 813 Santa Barbara St., Pasadena, CA 91101, USA}

\author[0000-0001-6533-6179]{Henrique M. Reggiani}
\affil{Gemini Observatory/NSF's NOIRLab, Casilla 603, La Serena, Chile}
\affil{Observatories of the Carnegie Institution for Science, 813 Santa Barbara St., Pasadena, CA 91101, USA}

\author[0000-0001-9261-8366]{Jhon Yana Galarza}
\affil{Observatories of the Carnegie Institution for Science, 813 Santa Barbara St., Pasadena, CA 91101, USA}


\correspondingauthor{T.~T.~Hansen}
\email{thidemannhansen@gmail.com}



\begin{abstract}
The growing number of Milky Way satellites detected in recent years has introduced a new focus for stellar abundance analysis. Abundances of stars in satellites have been used to probe the nature of these systems and their chemical evolution. However, for most satellites, only centrally located stars have been examined. This paper presents an analysis of three stars in the Tucana V system, one in the inner region and two at $\sim$10\arcmin\ (7--10 half-light radii) from the center. We find a remarkable chemical diversity between the stars. One star exhibits enhancements in rapid neutron-capture elements (an $r$-I star), and another is highly enhanced in C, N, and O but with low neutron-capture abundances (a CEMP-no star). The metallicities of the stars analyzed span more than 1~dex from $\mathrm{[Fe/H]}=-3.55$ to $-2.46$. This, combined with a large abundance range of other elements like Ca, Sc, and Ni, confirms that Tuc~V is an ultra-faint dwarf (UFD) galaxy. The variation in abundances, highlighted by [Mg/Ca] ratios ranging from $+0.89$ to $-0.75$, among the stars, demonstrates that the chemical enrichment history of Tuc~V was very inhomogeneous. Tuc~V is only the second UFD galaxy in which stars located at large distances from the galactic center have been analyzed, along with Tucana~II. The chemical diversity seen in these two galaxies, driven by the composition of the non-central member stars, suggests that distant member stars are important to include when classifying faint satellites and that these systems may have experienced more complex chemical enrichment histories than previously anticipated.
\end{abstract}


\section{Introduction \label{sec:intro}}

The population of Milky Way (MW) satellites has exploded over the last decade, especially at the faint end. Currently, 44 satellites that have luminosities $L \le 1000$~L$_{\odot}$ ($M_{V} \ge -2.7$) have been discovered, only eight of which were known prior to 2013. These stellar systems range in half-light radius from 1~pc to 90~pc, in distance from 5~kpc to 250~kpc, and have luminosities as faint as 45~L$_{\odot}$.

Determining the nature of such diminutive objects has proven to be a major challenge. Are they dark matter-dominated dwarf galaxies or dark matter-free star clusters? Did they reach their tiny sizes and luminosities via tidal stripping of larger systems, or did they form close to their current size?  Published spectroscopy to provide dynamical masses and chemical abundance information is available for only a small minority of this population. Consequently, in most cases, they have been classified based solely on their physical size, with satellites at $r_{\mathrm{half}} \lesssim 10$~pc regarded as clusters, those with $r_{\mathrm{half}} \gtrsim 30$~pc as dwarfs, and the nature of those in between remaining uncertain. In a few cases, the stellar kinematics and/or chemistry \citep{geha09,fadely11,simon11,kirby13,ji2019a,ji2020a,fu2023} have provided relatively confident classifications. Most recently, measurements of mass segregation from deep imaging have provided a promising route for separating dynamically relaxed clusters from dwarf galaxies that have not relaxed because of their large dark matter content \citep{kim15b,baumgardt22}.

A representative object in this category is Tucana~V (\tucv), which was discovered in imaging from the Dark Energy Survey by \citet{dw15}. \tucv\ has a luminosity of $240^{+170}_{-90}$~L$_{\odot}$, a half-light radius of $34^{+11}_{-8}$~pc (somewhat larger than the initially-estimated $17 \pm 6$~pc from \citealt{dw15}), and is located at a distance of $55^{+3}_{-8}$~kpc \citep{simon20}. Based on its diffuse appearance in deep imaging, \citet{conn18} suggested that \tucv\ is either an unbound part of the halo of the Small Magellanic Cloud (SMC) or a tidally disrupted star cluster. On the other hand, \citet{simon20} obtained medium-resolution spectroscopy of \tucv, identifying three stars as spectroscopic members of the system and showing that their velocities and proper motions are not consistent with those of the SMC.  Given this very small set of faint member stars, \citeauthor{simon20} were not able to place meaningful constraints on the velocity dispersion or metallicity dispersion of \tucv.  However, \citet{fu2023} tentatively detected a metallicity spread in \tucv\ of $\sigma_{\mathrm{[Fe/H]}} = 0.61^{+0.44}_{-0.28}$~dex using HST narrow-band imaging.

In this paper, we present a detailed chemical analysis of three stars in Tuc~V, one located in the center of the system, Gaia~DR3~6486878326228216320, hereafter Tuc~V-3, included in the sample of \citet{simon20}, and two newly detected members located at larger radii, Gaia~DR3~6486893242650962944 and Gaia~DR3~6486519060804598016 (hereafter Tuc~V-1 and Tuc~V-2, respectively). We further use this information to constrain the nature of Tuc~V and investigate the chemical enrichment of the system. The paper is organized as follows: observations and analysis of the stars are described in Sections \ref{sec:obs} and \ref{sec:analysis}, Section \ref{sec:results} presents our results, which are discussed in Section \ref{sec:discus}, and Section \ref{sec:summary} provides a summary.

\begin{figure*}[htb!]
\epsscale{1.17}
\plotone{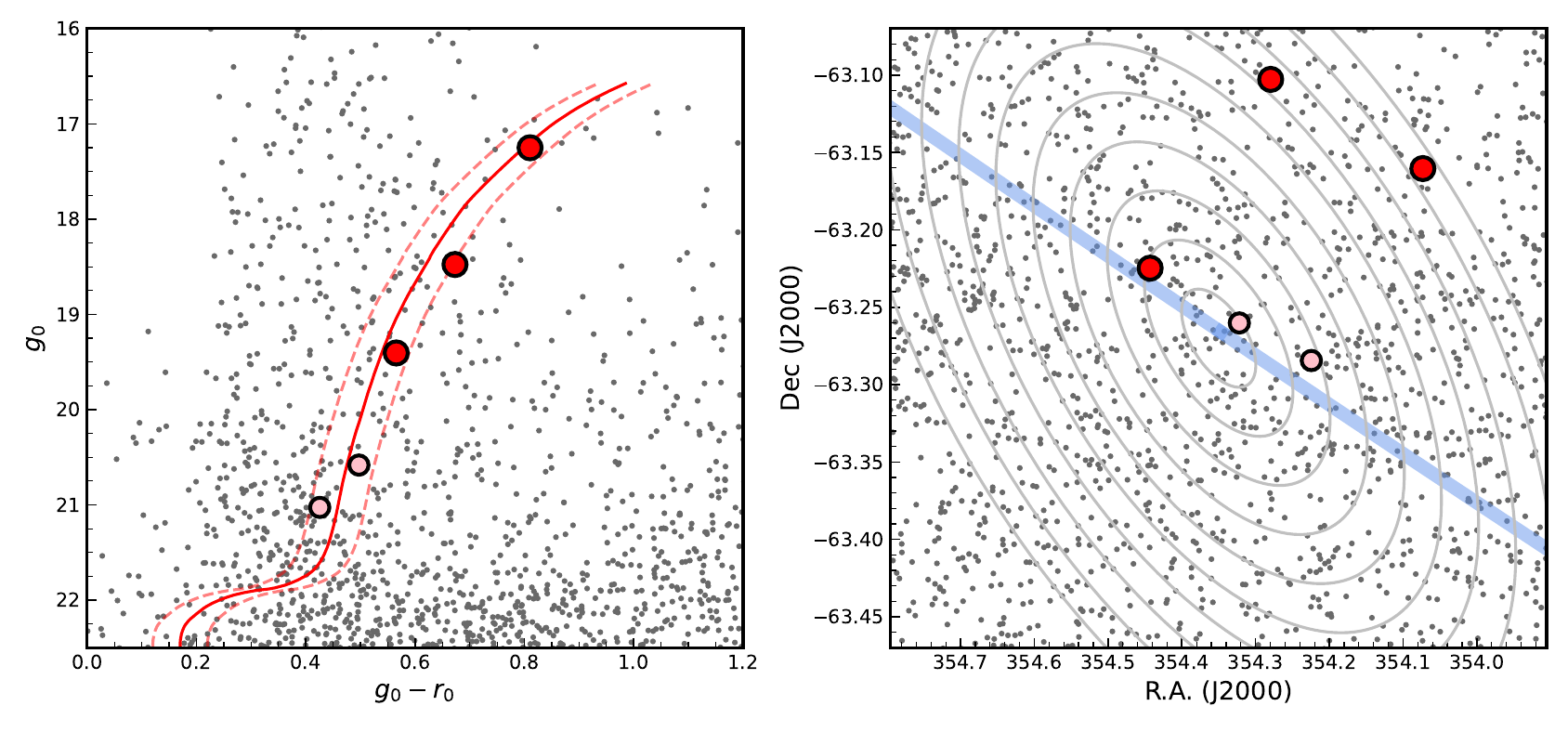}
\caption{(left) Color-magnitude diagram of a $24\arcmin \times 24\arcmin$ field centered on Tuc~V, using photometry from DES DR1 \citep{desdr1}. The three stars that are the subject of this paper are plotted as red circles, and the two fainter members of Tuc~V from \citet{simon20} are shown as smaller pink circles. A 12~Gyr, $\mathrm{[Fe/H]} = -2.49$ Dartmouth isochrone \citep{dotter2008} at a distance of 55~kpc is displayed in red, with additional curves (dashed lines) offset by $\pm0.05$~mag illustrating a typical photometric selection window around the isochrone. (right) Spatial distribution of stars around Tuc~V. The symbols are the same as in the left panel. Ellipses marking 1, 2, . . ., 10 times the half-light radius of Tuc~V are displayed in gray.  The orbital path of Tuc~V, computed with galpy \citep{galpy} based on the proper motion from \citet{pace22} and the updated radial velocity determined in \S~\ref{sec:nature}, is plotted as a thick blue line.}
\label{fig:cmd_spatial}
\end{figure*}

\begin{deluxetable*}{lcccccclrcccr}[htb!]
\tabletypesize{\footnotesize}
\tablecaption{\label{tab:obs} Observing log }
\tablehead{Gaia DR3 ID & Ra & Dec & $g_0$ & $r_0$ & $i_0$ & $z_0$ &MJD & $t_{\text{exp}}$ & SNR$^*$       & SNR$^*$ & $V_{\text{hel}}\pm\sigma$ &Instrument\\
              &    &     & (mag) & (mag) & (mag) & (mag) &    & (Sec)            & @4500\AA & @6500\AA & (kms$^{-1}$) & }
\startdata
6486893242650962944 & 23:37:07:09 & $-63$:06:10.2 & 17.25 & 16.44 & 16.13 & 15.99 & 59437.15 & 10800 & 22 & 51 & $-$23.9$\pm$0.6 & MIKE\\ 
(Tuc~V-1)           &             &               &       &       &       &       & 59552.03 & 900   &    &    & $-$42.0$\pm1.2$ & MIKE\\ 
                    &             &               &       &       &       &       & 59812.31 & 300   &    &    & $-$23.8$\pm$1.6 & MIKE\\  
                    &             &               &       &       &       &       & 59881.02 & 360   &    &    & $-$37.4$\pm$1.2 & IMACS\\ 
                    &             &               &       &       &       &       & 60095.41 & 900   &    &    & $-$26.1$\pm$1.0 & MIKE\\ 
                    &             &               &       &       &       &       & 60106.39 & 780   &    &    & $-$26.3$\pm$1.1 & IMACS\\ 
                    &             &               &       &       &       &       & 60113.27 & 1200  &    &    & $-$24.0$\pm$1.4 & MIKE\\ 
                    &             &               &       &       &       &       & 60177.43 & 2200  &    &    & $-$21.2$\pm$1.4 & MIKE\\ 
                    &             &               &       &       &       &       & 60250.02 & 600   &    &    & $-$33.7$\pm$1.7 & IMACS\\ 
                    &             &               &       &       &       &       & 60281.06 & 960   &    &    & $-$37.7$\pm$1.1 & MIKE\\ 
                    &             &               &       &       &       &       & 60306.06 & 960   &    &    & $-$41.8$\pm$1.3 & MIKE\\ 
6486519060804598016 & 23:36:17:58 & $-63$:09:37.6 & 18.48 & 17.80 & 17.56 & 17.43 & 59437.23 & 19900 & 19 & 44 & $-$33.8$\pm$0.6 & MIKE\\ 
(Tuc~V-2)           &             &               &       &       &       &       & 59813.13 & 11700 &    &    & $-$34.5$\pm$0.6 & MIKE\\  
6486878326228216320 & 23:37:46.33 & $-63$:13:29.2 & 19.41 & 18.84 & 18.63 & 18.54 & 59437.39 & 9900  & 15 & 33 & $-$35.4$\pm$1.6 & MIKE\\   
(Tuc~V-3)           &             &               &       &       &       &       & 59551.17 & 10800 &    &    & $-$36.1$\pm$0.9 & MIKE\\   
                    &             &               &       &       &       &       & 59552.14 & 8400  &    &    & $-$36.4$\pm$1.7 & MIKE\\   
                    &             &               &       &       &       &       & 59813.31 & 18120 &    &    & $-$36.0$\pm$0.9 & MIKE 
\enddata
\tablecomments{$^*$ of co-added MIKE spectra used for abundance analysis.}
\end{deluxetable*}

\section{Observations \label{sec:obs}}

\subsection{Target Selection}

The three \tucv\ member stars identified by \citet{simon20} have magnitudes of $g=19.42$, $g=20.60$, and $g=21.06$.  The brightest of these (Tuc~V-3) is accessible for high-resolution spectroscopy with long exposure times \citep[c.f.][]{simon2010,frebel2014}, but the other two stars are too faint for chemical abundance analysis from high-resolution spectra with current instruments.  However, we used astrometry from Gaia Early Data Release 3 \citep{gaiamission,gaiarv21} to identify two substantially brighter stars with proper motions matching that of the brightest \citeauthor{simon20} member (Tuc~V-1 and Tuc~V-2). These stars, at $G=16.57$ and $G=17.92$, are both located $\sim10\arcmin$ ($\sim5r_{\rm half}$, although their projected distances are larger because they are located close to the minor axis) from the center of Tuc~V and have colors and magnitudes that are consistent with the Tuc~V red giant branch (see Figure~\ref{fig:cmd_spatial}).  Although it would be surprising for the two brightest stars in the system to be located at such large radii if Tuc~V has an extended envelope similar to ultra-faint dwarf (UFD) galaxies such as Tucana~II \citep{chiti2021} or Bo{\"o}tes~I \citep{filion21,longeard22} then these stars could plausibly be Tuc~V members.  In order to test this possibility, we obtained high-resolution spectroscopy of the brightest confirmed Tuc~V member, as well as the two new candidates.

\subsection{Magellan/MIKE Spectroscopy}

The high-resolution spectra of the three stars used for abundance analysis were obtained with the MIKE spectrograph \citep{bernstein2003,shectman2003} on the Magellan/Clay Telescope at Las Campanas Observatory in 2021 August, 2021 December, and 2022 August. Additional low signal-to-noise (SNR) MIKE spectra of Tuc~V-1 for radial velocity measurements were also obtained through 2023 August. Table \ref{tab:obs} lists target Gaia DR3 IDs, R.A., Decl., DES DR1 magnitudes, modified Julian date (MJD) of the observations, exposure times, SNR, and heliocentric velocities for the spectra. The radial velocities are determined via $\chi^{2}$ fits of the spectra with a template spectrum of the radial-velocity standard star HD122563 ($V_{hel}=-26.17$~km~s$^{-1}$; \citealt{gaiarv21}) also obtained with the MIKE spectrograph, as described by \citet{ji2020a}. The MIKE spectra cover wavelength ranges of 3330\AA$<\lambda<$5060\AA\ in the blue channel and 4830\AA$<\lambda<$9410\AA\ in the red. For Tuc~V-3, the $1\arcsec\times5\arcsec$ slit and $2\times2$ binning were used, yielding a typical resolution of $R\sim$ 28000/22000 in the blue and red arms, and for the other two stars, the $0\farcs7\times5\arcsec$ slit and $2\times2$ binning was used, yielding a typical resolution of $R\sim$ 35000/28000 in the blue and red arms. All data were reduced with the CarPy MIKE pipeline \citep{kelson2000,kelson2003}, and multiple spectra of the same star from different runs were subsequently co-added. The seeing on the observing nights listed in Table \ref{tab:obs} varied from 0.5" to 1.4", but for most of the observations, the seeing was not significantly smaller than the slit width. Thus, slit centering has not resulted in any meaningful offsets on the measured radial velocities as confirmed by the good agreement between individual measurements of Tuc~V-2 and Tuc~V-3.

\subsection{Magellan/IMACS Spectroscopy}

We obtained two additional spectra of Tuc~V-1 in 2022 October and 2023 June with the IMACS spectrograph \citep{dressler2011} on the Magellan/Baade telescope to measure its radial velocity.  For these observations, we employed a 0\farcs7-wide long slit with the 1200-line grating on the $f/4$ camera, resulting in a resolution of $R\sim 11,000$.  Data reduction and analysis procedures followed those described by \citet{simon2017} and \citet{li2017} and subsequent papers.

\subsection{Tuc~V Membership}

Using the initial MIKE spectrum of Tuc~V-2, we measured a velocity of  $v_{hel} = -33.8 \pm 0.6$~km~s$^{-1}$ and a metallicity of $\mbox{[Fe/H]} < -2$.  This velocity agrees within the uncertainties with the systemic velocity of \tucv\ ($-36.2^{+2.5}_{-2.2}$~km~s$^{-1}$) \citep{simon20}, and combined with the low metallicity, confirms that the star belongs to \tucv.  In contrast, our first velocity measurement for Tuc~V-1 was offset from \tucv\ by 12~km~s$^{-1}$.  However, the extremely low metallicity of this star makes it very unlikely to be a foreground halo star unrelated to \tucv.  We, therefore, suspected that the star is a member of \tucv\ with a binary companion and began monitoring its radial velocity.  The subsequent measurements listed in Table~\ref{tab:obs} confirmed this hypothesis, revealing periodic orbital motion with a center-of-mass velocity close to that of the other \tucv\ stars.  It is thus safe to conclude that both bright stars are \tucv\ members.  Velocity measurements of Tuc~V-2 and Tuc~V-3 spanning 1~yr for each star show no evidence of velocity variability, indicating that neither star is likely to be in a binary system with a period shorter than several years.

As mentioned above, these two new \tucv\ stars are located at large distances from the center of the system.  If \tucv\ has a normal exponential or Plummer radial profile, it is very unlikely that two of its handful of red giants would be found at $\gtrsim7r_{\rm half}$.  This result, therefore, suggests that either the size of \tucv\ has been significantly underestimated or that it has an extended stellar component with a different radial profile.  In the latter case, the \tucv\ halo could have originated in an ancient merger \citep[e.g.,][]{chiti2021,tarumi2021,goater2023} or could indicate that \tucv\ is undergoing tidal stripping by the Milky Way.  We encourage deeper and wider imaging of \tucv\ to explore these possibilities further.

\section{Abundance Analysis \label{sec:analysis}} 
The spectral analysis was done using the analysis code \code{SMHR}\footnote{\href{https://github.com/andycasey/smhr}{https://github.com/andycasey/smhr}} to run the 2017 version of the radiative transfer code \code{MOOG}\footnote{\href{https://github.com/alexji/moog17scat} {https://github.com/alexji/moog17scat}} \citep{sneden1973,sobeck2011}. The analysis assumes local thermodynamical equilibrium and one-dimensional $\alpha$-enhanced ($\mathrm{[\alpha/Fe] = +0.4}$) \code{ATLAS9} model atmospheres \citep{castelli2003} were used as input. The line lists used were generated from \code{linemake}\footnote{\href{https://github.com/vmplacco/linemake}{https://github.com/vmplacco/linemake}} \citep{placco2021} and include isotopic (IS) and hyperfine structure (HFS) broadening, where applicable, employing the $r$-process isotope ratios from \citet{sneden2008}. Finally, Solar abundances were taken from \citet{asplund2009}.

The stellar parameters were determined from a combination of photometry and equivalent width (EW) measurements of \ion{Fe}{1} and \ion{Fe}{2} lines obtained by fitting Gaussian profiles to the absorption lines in the continuum-normalized spectra. Photometric effective temperatures ($T_{\rm eff}$) were determined from de-reddened $g$, $r$, $i$, and $z$ colors from DES DR1 \citep{desdr1}, listed in Table \ref{tab:obs}, converted to the corresponding $B$, $V$, $R$, and $I$ colors \citep[][ R. Lupton 2005\footnote{\href{http://www.sdss3.org/dr8/algorithms/sdssUBVRITransform.php}{http://www.sdss3.org/dr8/algorithms/sdssUBVRITransform.php}}]{drlicawagner2018}, and employing the $B-V$, $V-R$, $R-I$, and $V-I$ color temperature relations from \citet{casagrande2010}. The final temperature is taken as an average of the four indicators. Individual temperatures from the four colors are listed in Table \ref{tab:teff}, along with spectroscopic temperatures for the stars that have been placed on a photometric scale using the relation from \citet{frebel2013}. The photometric and corrected spectroscopic temperatures agree within uncertainties. In the top panel of Figure \ref{fig:param}, we plot the \ion{Fe}{1} and \ion{Fe}{2} line abundances as a function of excitation potential ($\chi$). A small trend is visible in these plots due to the offset between the spectroscopic and photometric temperature scales. Following the determination of $T_{\rm eff}$, the surface gravities ($\log g$) were determined from ionization equilibrium between the \ion{Fe}{1} and \ion{Fe}{2} lines and microturbulent velocities ($\xi$) by removing any trend in line abundances with reduced EW for the \ion{Fe}{1} lines (see Figure \ref{fig:param}). The model atmosphere metallicity ($\mathrm{[M/H]}$) is taken as the mean of the \ion{Fe}{1} and \ion{Fe}{2} abundances. Final stellar parameters and their uncertainties are presented in Table \ref{tab:params}. The uncertainty on $T_{\rm eff}$ is taken as the standard deviation of $T_{\rm eff}$ from the four colors used, and the effect of the standard deviation of abundances of \ion{Fe}{1} lines used added in quadrature. Uncertainties on $\log g$, $\mathrm{[M/H]}$, and $\xi$ are the result of the photometric uncertainty on $T_{\rm eff}$ and the standard deviation of abundances of \ion{Fe}{1} lines used added in quadrature. The effects of the standard deviation of the \ion{Fe}{1} line abundances were determined by varying each parameter ($T_{\rm eff}$, $\log g$, and $\xi$) one at a time, to match the standard deviation of \ion{Fe}{1} line abundances. Individual stellar parameter uncertainties are listed in Table \ref{tab:paramerr}.

Following the stellar parameter determination, elemental abundances were derived from EW measurements and spectral synthesis. Table \ref{tab:lines} lists the atomic data for the lines used in the analysis and the measured EW and abundance of each line. Final mean abundances and associated uncertainties were determined following the procedure outlined in \cite{ji2020b}. This method performs a line-by-line analysis and includes both statistical and systematic uncertainties and covariance terms between stellar parameter uncertainties. 


\begin{deluxetable}{lcccc}[htb!]
\caption{Stellar Parameters\label{tab:params}}
\tablehead{ID & $T_{\rm eff}$ & log $g$ & $\xi$  & $\mathrm{[Fe/H]}$ \\
&(K)&(cgs)&(km~s$^{-1}$)& }
\startdata
Tuc~V-1 & 4380$\pm104$ & 0.54$\pm0.32$ & 2.18$\pm0.09$ & -3.47$\pm$0.21\\ 
Tuc~V-2 & 4773$\pm92$  & 1.53$\pm0.21$ & 1.91$\pm0.08$ & -2.41$\pm$0.15 \\ 
Tuc~V-3 & 4982$\pm130$ & 1.70$\pm0.20$ & 1.61$\pm0.13$ & -2.57$\pm$0.25 \\ 
\enddata
\end{deluxetable}

\begin{deluxetable}{crrrrrrrrrrrrc}
\tablecaption{\label{tab:lines} Data for atomic lines used in analysis and individual line EW measurements for each star. Atomic numbers are listed under species.}
\tablehead{
\colhead{ID} &\colhead{Species} & \colhead{$\lambda$} & \colhead{$\chi$} & \colhead{$\log gf$} & \colhead{EW} &  \colhead{$\sigma_{\rm EW}$} & \colhead{$\log\epsilon$} & \colhead{ref}\\
\colhead{} & \colhead{} &\colhead{(\AA)} & \colhead{(eV)} & \colhead{} & \colhead{(m\AA)} & \colhead{(m\AA)} & \colhead{} & \colhead{} }
\startdata
Tuc V-1 &   8.0 & 6300.30 & 0.00 & $-$9.69 &    31.01 &   4.67 & $+$6.92 &  1\\
Tuc V-1 &   8.0 & 7771.94 & 9.15 & $+$0.37 &    11.01 &   4.47 & $+$7.46 &  1\\
Tuc V-1 &   8.0 & 7775.39 & 9.15 &    0.00 &    14.01 &   3.10 & $+$7.97 &  1\\
Tuc V-1 &  11.0 & 5889.95 & 0.00 & $+$0.11 &   206.25 &   3.80 & $+$3.94 &  1\\
Tuc V-1 &  11.0 & 5895.92 & 0.00 & $-$0.19 &   180.34 &   3.60 & $+$3.90 &  1\\
Tuc V-1 &  11.0 & 8183.26 & 2.10 & $+$0.24 &    25.62 &   2.91 & $+$3.49 &  1\\
Tuc V-1 &  11.0 & 8194.81 & 2.10 & $+$0.49 &    48.89 &   3.22 & $+$3.64 &  1\\
Tuc V-1 &  12.0 & 4057.51 & 4.35 & $-$0.90 &    63.95 &   6.38 & $+$5.52 &  1\\

\enddata
\tablerefs{(1) \citet{kramida2021}; (2) \citet{pehlivan2017}; (3) \citet{denhartog2023}; (4) \citet{denhartog2021}; (5) \citet{lawler2019}; (6) \citet{lawler2013}; (7) \citet{wood2013}; (8) \citet{pickering2001}, with corrections given in \citet{pickering2002}; (9) \citet{wood2014a} for log(gf) values and HFS, when available; (10) \citet{sobeck2007}; (11) \citet{lawler2017}; (12) \citet{denhartog2011} for both log(gf) value and HFS; (13) \citet{obrian1991}; (14) \citet{ruffoni2014}; (15) \citet{belmonte2017}; (16) \citet{denhartog2014}; (17) \citet{blackwell1982}; (18) \citet{denhartog2019}; (19) \citet{melendez2009}; (20) \citet{lawler2015} for log(gf) values and HFS; (21) \citet{wood2014b}; (22) \citet{roederer2012}; (23) \citet{biemont2011}; (24) \citet{ljung2006}; (25) \citet{kramida2021}, using HFS/IS from \citet{mcwilliam1998} when available; (26) \citet{lawler2001a}, using HFS from \citet{ivans2006} when available; (27) \citet{lawler2009}; (28) \citet{li2007}, using HFS from \citet{sneden2009}; (29) \citet{ivarsson2001}, using HFS from \citet{sneden2009}; (30) \citet{lawler2001b}, using HFS/IS from \citet{ivans2006}; (31) \citet{wickliffe2000}}
\tablecomments{The complete version of this Table is available online only. A short version is shown here to illustrate its form and content.}
\end{deluxetable}

\section{Abundance Results \label{sec:results}}
\begin{figure*}[hbt!]
\centering
\includegraphics[width=\linewidth]{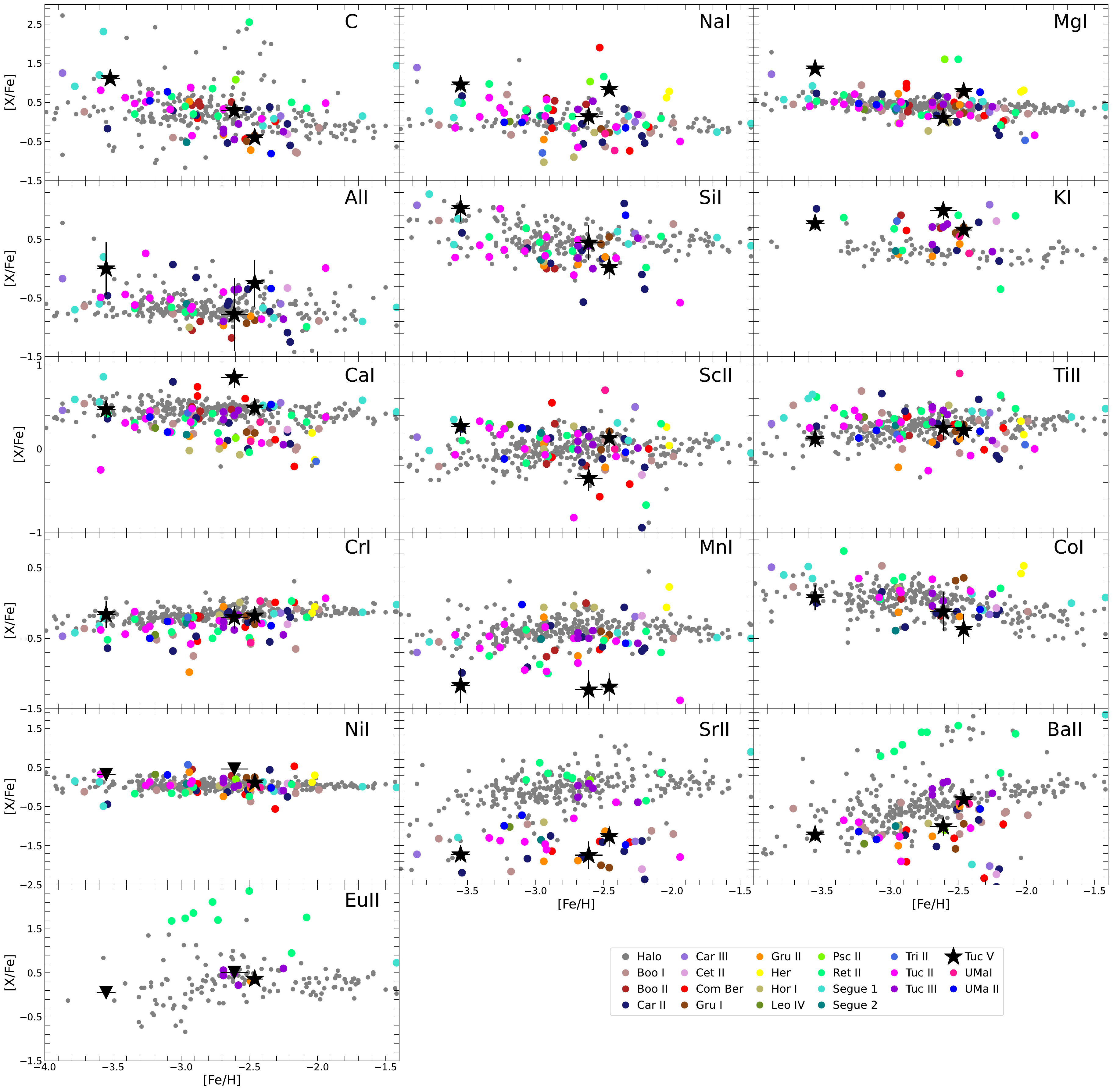}
\caption{\label{fig:abun} $\mathrm{[X/Fe]}$ derived abundances for Tuc~V (black stars) compared to stellar abundances from the MW halo \citep[grey dots;][]{roederer2014a} and other UFD galaxies (colored dots according to legend, see text for references). Upper limits for Tuc~V stars are marked with downward pointing black triangles}
\end{figure*}

Our final abundances and total associated uncertainties for the three stars are listed in Table~\ref{tab:abun}. Abundance uncertainties arising from individual stellar parameters ($\Delta_{\rm X}$) are listed in Table \ref{tab:uncer} along with the systematic uncertainty $s_X$. We were able to derive abundances for 15 elements from C to Ba in all three stars and additional elements in Tuc~V-2 and Tuc~V-1 (see details below). Abundances of elements from C to Eu of the three stars are shown in Figure \ref{fig:abun} compared to abundance results from high-resolution ($R\gtrsim 25,000$) studies of other UFD galaxies and metal-poor MW halo stars \citep{roederer2014}. Only abundance detections have been included in the comparison sample. The UFD galaxies with literature abundance measurements are: Bo\"{o}tes~I \citep{feltzing2009,frebel2016,gilmore2013,ishigaki2014,norris2010,waller2023}, Bo\"{o}tes~II \citep{ji2016a}, Carina~II \citep{ji2020a}, Carina~III \citep{ji2020a}, Cetus~II \citep{webber2023}, Coma Berenices \citep{frebel2010,waller2023}, Grus~I \citep{ji2019a}, Grus~II \citep{hansen2020}, Hercules \citep{koch2008}, Horologium~I \citep{nagasawa2018}, Leo~IV \citep{simon2010}, Pisces~II \citep{spite2018}, Reticulum~II \citep{ji2016b,ji2016c,hayes2023}, Segue~1 \citep{frebel2014,norris2010}, Segue~2 \citep{roederer2014a}, Triangulum~II \citep{ji2019a}, Tucana~II \citep{ji2016d,chiti2018,chiti2023}, Tucana~III \citep{hansen2017,marshall2019}, Ursa Major~I \citep{waller2023}, and Ursa Major~II \citep{frebel2010}.

\begin{deluxetable*}{lrrrrrrrrrrrrrrrrrr}[htb!]
\centerwidetable
\tablecolumns{12}
\tablewidth{\textwidth}
\tabletypesize{\scriptsize}
\tablecaption{\label{tab:abun}Abundance Summary for the three Tuc~V stars}
\tablehead{\multicolumn{1}{c}{}&\multicolumn{6}{c}{Tuc~V-1} & \multicolumn{6}{c}{Tuc~V-2} & \multicolumn{6}{c}{Tuc~V-3} \\
 \colhead{species} & \colhead{N} & \colhead{$\log\epsilon$ (X)} & \colhead{$\mathrm{[X/H]}$} & \colhead{$\sigma_{\mathrm{[X/H]}}$} & \colhead{$\mathrm{[X/Fe]}$} & \colhead{$\sigma_{\mathrm{[X/Fe]}}$} & \colhead{N} & \colhead{$\log\epsilon$ (X)} & \colhead{$\mathrm{[X/H]}$} & \colhead{$\sigma_{\mathrm{[X/H]}}$} & \colhead{$\mathrm{[X/Fe]}$} & \colhead{$\sigma_{\mathrm{[X/Fe]}}$} & \colhead{N} & \colhead{$\log\epsilon$ (X)} & \colhead{$\mathrm{[X/H]}$} & \colhead{$\sigma_{\mathrm{[X/H]}}$} & \colhead{$\mathrm{[X/Fe]}$} & \colhead{$\sigma_{\mathrm{[X/Fe]}}$}\\
 &  &  &  & [dex] & & [dex] &  &  &  & [dex] &  & [dex] &  &  &  & [dex] &  & [dex]}
\startdata
  C-H       &    &  $+$5.99 & $-$2.44 & 0.13  & $+$1.11 & 0.11  &    & $+$5.56 & $-$2.87 & 0.10  & $-$0.40 & 0.09  &   & $+$6.10 & $-$2.33 & 0.17  & $+$0.29 & 0.15  \\
  C$_{corr}$&    &  $+$6.65 & $-$1.78 & 0.13  & $+$1.77 & 0.11  &    & $+$6.06 & $-$2.37 & 0.10  & $+$0.10 & 0.09  &   & $+$6.37 & $-$2.05 & 0.17  & $+$0.56 & 0.15  \\
  C-N       &    &  $+$6.09 & $-$1.75 & 0.35  & $+$1.81 & 0.34  &    &\nodata  &\nodata  &\nodata&\nodata  &\nodata&   &\nodata  &\nodata  &\nodata& \nodata &\nodata\\
\ion{O}{1}  &  3 &  $+$7.46 & $-$1.23 & 0.26  & $+$2.32 & 0.27  &    &\nodata  &\nodata  &\nodata&\nodata  &\nodata&   &\nodata  &\nodata  &\nodata& \nodata &\nodata\\
\ion{Na}{1} &  4 &  $+$3.64 & $-$2.60 & 0.15  & $+$0.95 & 0.15  &  5 & $+$4.62 & $-$1.62 & 0.10  & $+$0.84 & 0.10  & 2 & $+$3.77 & $-$2.47 & 0.23  & $+$0.14 & 0.21  \\
\ion{Mg}{1} &  6 &  $+$5.41 & $-$2.19 & 0.10  & $+$1.36 & 0.11  &  8 & $+$5.92 & $-$1.68 & 0.09  & $+$0.78 & 0.10  & 5 & $+$5.09 & $-$2.51 & 0.11  & $+$0.10 & 0.12  \\
\ion{Al}{1} &  1 &  $+$2.90 & $-$3.55 & 0.45  &    0.00 & 0.45  &  1 & $+$3.73 & $-$2.72 & 0.41  & $-$0.25 & 0.40  & 1 & $+$3.06 & $-$3.39 & 0.62  & $-$0.78 & 0.62  \\
\ion{Si}{1} &  1 &  $+$4.99 & $-$2.52 & 0.22  & $+$1.03 & 0.23  &  2 & $+$5.07 & $-$2.44 & 0.20  & $+$0.02 & 0.19  & 2 & $+$5.34 & $-$2.17 & 0.32  & $+$0.44 & 0.30  \\
\ion{K}{1}  &  2 &  $+$2.25 & $-$2.78 & 0.11  & $+$0.77 & 0.11  &  2 & $+$3.23 & $-$1.80 & 0.12  & $+$0.66 & 0.11  & 1 & $+$3.41 & $-$1.62 & 0.16  & $+$0.99 & 0.15  \\
\ion{Ca}{1} & 16 &  $+$3.26 & $-$3.08 & 0.09  & $+$0.47 & 0.09  & 15 & $+$4.37 & $-$1.97 & 0.08  & $+$0.49 & 0.08  &15 & $+$4.58 & $-$1.76 & 0.11  & $+$0.85 & 0.12  \\
\ion{Sc}{2} &  7 &  $-$0.02 & $-$3.17 & 0.14  & $+$0.27 & 0.12  &  7 & $+$0.84 & $-$2.31 & 0.12  & $+$0.13 & 0.10  & 5 & $+$0.20 & $-$2.96 & 0.19  & $-$0.35 & 0.15  \\
\ion{Ti}{1} &  4 &  $+$1.70 & $-$3.25 & 0.16  & $+$0.30 & 0.15  & 14 & $+$2.71 & $-$2.24 & 0.12  & $+$0.22 & 0.11  & 2 & $+$2.47 & $-$2.48 & 0.20  & $+$0.13 & 0.19  \\
\ion{Ti}{2} & 16 &  $+$1.63 & $-$3.32 & 0.15  & $+$0.12 & 0.10  & 16 & $+$2.73 & $-$2.22 & 0.13  & $+$0.22 & 0.11  &12 & $+$2.58 & $-$2.37 & 0.15  & $+$0.25 & 0.14  \\
\ion{V}{2}  &    & $<+$1.11 &$<-$2.82 &\nodata&$<+$0.62 &\nodata&  1 & $+$1.58 & $-$2.35 & 0.16  & $+$0.08 & 0.16  &   &$<+$2.05 &$<-$1.88 &\nodata&$<+$0.69 &\nodata\\
\ion{Cr}{1} &  2 &  $+$1.93 & $-$3.71 & 0.19  & $-$0.16 & 0.19  & 10 & $+$3.00 & $-$2.64 & 0.12  & $-$0.18 & 0.11  & 6 & $+$2.83 & $-$2.81 & 0.17  & $-$0.20 & 0.16  \\
\ion{Cr}{2} &    & \nodata  & \nodata &\nodata& \nodata &\nodata&  2 & $+$3.08 & $-$2.56 & 0.17  & $-$0.12 & 0.16  &   & \nodata & \nodata &\nodata& \nodata &\nodata\\
\ion{Mn}{1} &  2 &  $+$0.71 & $-$4.72 & 0.26  & $-$1.17 & 0.25  &  3 & $+$1.78 & $-$3.65 & 0.21  & $-$1.19 & 0.20  & 3 & $+$1.58 & $-$3.85 & 0.28  & $-$1.23 & 0.28  \\
\ion{Fe}{1} & 68 &  $+$3.95 & $-$3.55 & 0.08  & \nodata &\nodata&112 & $+$5.04 & $-$2.46 & 0.06  & \nodata &\nodata&65 & $+$4.89 & $-$2.61 & 0.10  & \nodata &\nodata\\
\ion{Fe}{2} &  4 &  $+$4.06 & $-$3.44 & 0.13  & \nodata &\nodata& 11 & $+$5.06 & $-$2.44 & 0.10  & \nodata &\nodata& 6 & $+$4.89 & $-$2.61 & 0.14  & \nodata &\nodata\\
\ion{Co}{1} &  3 &  $+$1.55 & $-$3.48 & 0.20  & $+$0.08 & 0.18  &  3 & $+$2.16 & $-$2.83 & 0.22  & $-$0.37 & 0.21  & 2 & $+$2.26 & $-$2.73 & 0.29  & $-$0.12 & 0.28  \\
\ion{Ni}{1} &    & $<+$3.09 &$<-$3.10 &\nodata&$<+$0.32 &\nodata&  6 & $+$3.86 & $-$2.36 & 0.11  & $+$0.11 & 0.10  &   &$<+$4.11 &$<-$2.11 &\nodata&$<+$0.46 &\nodata\\
\ion{Zn}{1} &    & $<+$1.92 &$<-$2.64 &\nodata&$<+$0.80 &\nodata&  2 & $+$2.47 & $-$2.09 & 0.13  & $+$0.37 & 0.13  &   &$<+$2.54 &$<-$2.02 &\nodata&$<+$0.55 &\nodata\\
\ion{Sr}{2} &  2 &  $-$2.30 & $-$5.17 & 0.18  & $-$1.73 & 0.16  &  2 & $-$0.82 & $-$3.69 & 0.31  & $-$1.26 & 0.27  & 1 & $-$1.48 & $-$4.35 & 0.38  & $-$1.74 & 0.35  \\
\ion{Y}{2}  &    &  \nodata & \nodata &\nodata& \nodata &\nodata&  3 & $-$0.91 & $-$3.12 & 0.22  & $-$0.69 & 0.21  &   & \nodata & \nodata &\nodata& \nodata &\nodata\\
\ion{Zr}{2} &    &  \nodata & \nodata &\nodata& \nodata &\nodata&  1 & $-$0.50 & $-$3.08 & 0.35  & $-$0.65 & 0.35  &   & \nodata & \nodata &\nodata& \nodata &\nodata\\
\ion{Ba}{2} &  4 &  $-$2.48 & $-$4.66 & 0.20  & $-$1.22 & 0.16  &  5 & $-$0.58 & $-$2.76 & 0.15  & $-$0.32 & 0.11  & 2 & $-$1.44 & $-$3.62 & 0.27  & $-$1.01 & 0.23  \\
\ion{La}{2} &    &  \nodata & \nodata &\nodata& \nodata &\nodata&  2 & $-$1.55 & $-$2.65 & 0.20  & $-$0.22 & 0.19  &   & \nodata & \nodata &\nodata& \nodata &\nodata\\
\ion{Ce}{2} &    &  \nodata & \nodata &\nodata& \nodata &\nodata&  2 & $-$0.81 & $-$2.39 & 0.31  & $+$0.04 & 0.32  &   & \nodata & \nodata &\nodata& \nodata &\nodata\\
\ion{Pr}{2} &    &  \nodata & \nodata &\nodata& \nodata &\nodata&  1 & $-$1.34 & $-$2.06 & 0.30  & $+$0.37 & 0.27  &   & \nodata & \nodata &\nodata& \nodata &\nodata\\
\ion{Nd}{2} &    &  \nodata & \nodata &\nodata& \nodata &\nodata&  5 & $-$0.58 & $-$2.00 & 0.13  & $+$0.43 & 0.12  &   & \nodata & \nodata &\nodata& \nodata &\nodata\\
\ion{Eu}{2} &    & $<-$2.87 &$<-$3.39 &\nodata&$<+$0.05 &\nodata&  2 & $-$1.53 & $-$2.05 & 0.14  & $+$0.36 & 0.11  &   &$<-$1.54 &$<-$2.06 &\nodata&$<+$0.51 &\nodata\\
\ion{Dy}{2} &    &  \nodata & \nodata &\nodata& \nodata &\nodata&  1 & $-$1.16 & $-$2.26 & 0.35  & $+$0.18 & 0.36  &   & \nodata & \nodata &\nodata& \nodata &\nodata

\enddata
\end{deluxetable*}

\begin{deluxetable}{llrrrrr}[htb!]
\centerwidetable
\tablecolumns{7}
\tablecaption{\label{tab:uncer} Abundance Uncertainties}
\tablehead{\colhead{ID} & \colhead{El.} & \colhead{$\Delta_{T_{\rm eff}}$} & \colhead{$\Delta_{\log g}$} & \colhead{$\Delta_{\xi}$} & \colhead{$\Delta_\mathrm{[Fe/H]}$} & \colhead{$s_X$}  \\
 & & [dex] & [dex] & [dex] & [dex] & [dex]}
\startdata
Tuc V-1 &   C-H        &    0.10 &  $-$0.07 & $-$0.01 &    0.03 &  0.07 \\
Tuc V-1 &   C-N        &    0.20 &  $-$0.02 &    0.00 &    0.04 &  0.00 \\
Tuc V-1 & \ion{O}{1}   & $-$0.07 &     0.10 & $-$0.01 &    0.01 &  0.43 \\
Tuc V-1 & \ion{Na}{1}  &    0.09 &  $-$0.05 & $-$0.03 & $-$0.03 &  0.21 \\
Tuc V-1 & \ion{Mg}{1}  &    0.08 &  $-$0.10 & $-$0.06 & $-$0.03 &  0.17 \\
Tuc V-1 & \ion{Al}{1}  &    0.15 &  $-$0.13 & $-$0.13 & $-$0.08 &  0.30 \\
Tuc V-1 & \ion{Si}{1}  & $-$0.02 &  $-$0.06 & $-$0.07 & $-$0.05 &  0.00 \\
Tuc V-1 & \ion{K}{1}   &    0.10 &  $-$0.03 & $-$0.01 & $-$0.02 &  0.00 \\
Tuc V-1 & \ion{Ca}{1}  &    0.09 &  $-$0.04 & $-$0.01 & $-$0.02 &  0.09 \\
Tuc V-1 & \ion{Sc}{2}  & $-$0.01 &     0.08 & $-$0.04 &    0.00 &  0.20 \\
Tuc V-1 & \ion{Ti}{1}  &    0.16 &  $-$0.05 & $-$0.01 & $-$0.02 &  0.10 \\
Tuc V-1 & \ion{Ti}{2}  &    0.04 &     0.07 & $-$0.03 &    0.02 &  0.13 \\
Tuc V-1 & \ion{Cr}{1}  &    0.14 &  $-$0.05 & $-$0.02 & $-$0.03 &  0.19 \\
Tuc V-1 & \ion{Mn}{1}  &    0.23 &  $-$0.10 & $-$0.05 & $-$0.04 &  0.15 \\
Tuc V-1 & \ion{Fe}{1}  &    0.08 &  $-$0.02 &    0.03 &    0.01 &  0.21 \\
Tuc V-1 & \ion{Fe}{2}  &    0.00 &     0.09 & $-$0.02 &    0.02 &  0.11 \\
Tuc V-1 & \ion{Co}{1}  &    0.16 &  $-$0.06 & $-$0.05 & $-$0.04 &  0.09 \\
Tuc V-1 & \ion{Sr}{2}  & $-$0.02 &     0.07 & $-$0.07 & $-$0.01 &  0.00 \\
Tuc V-1 & \ion{Ba}{2}  &    0.06 &     0.08 & $-$0.00 &    0.03 &  0.19 \\

\enddata
\tablecomments{The complete version of this Table is available online only. A short version is shown here to illustrate its form and content.}
\end{deluxetable}

\subsection{CNO and odd Z elements}
\begin{figure}[hbt!]
\centering
\includegraphics[width=\linewidth]{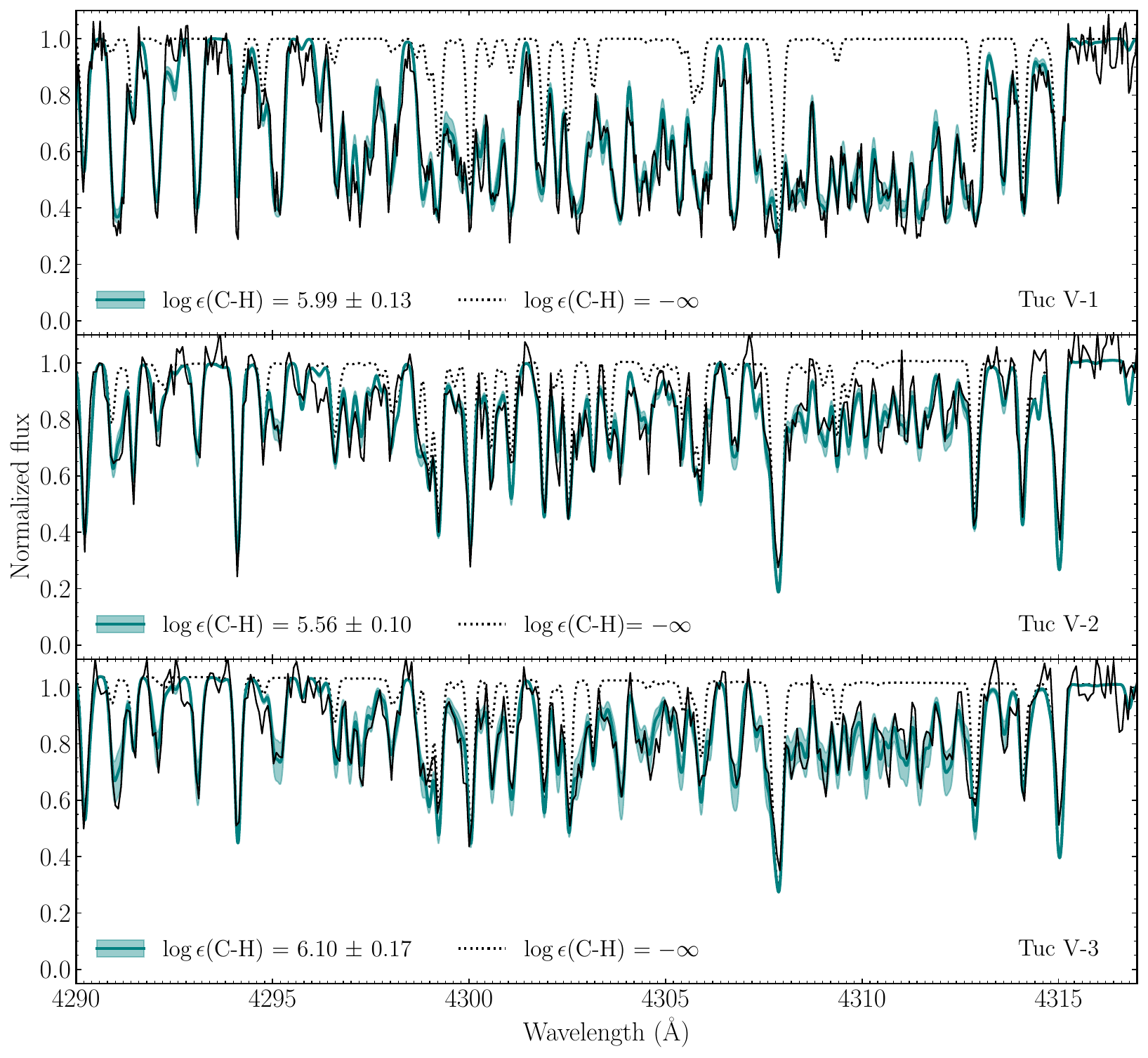}
\caption{\label{fig:CH}  Comparison of synthesis and observed spectra (black line) for the C-H G-band in Tuc~V-1 (top), Tuc~V-2 (middle), and Tuc~V-3 (bottom). The blue line is the best-fit synthesis, the blue band shows the uncertainty, and the dotted line is a synthesis without the given element.} 
\end{figure}

\begin{figure*}[hbt!]
\centering
\includegraphics[width=0.32\linewidth]{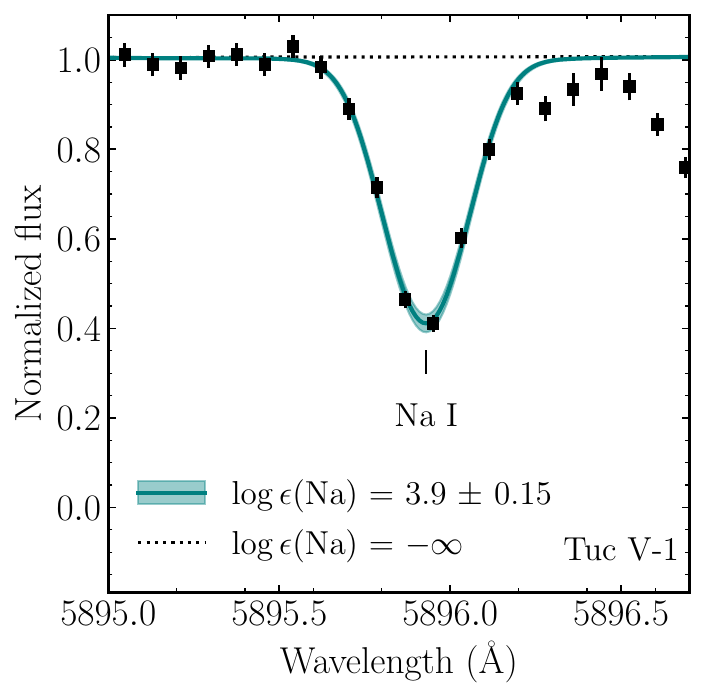}
\includegraphics[width=0.32\linewidth]{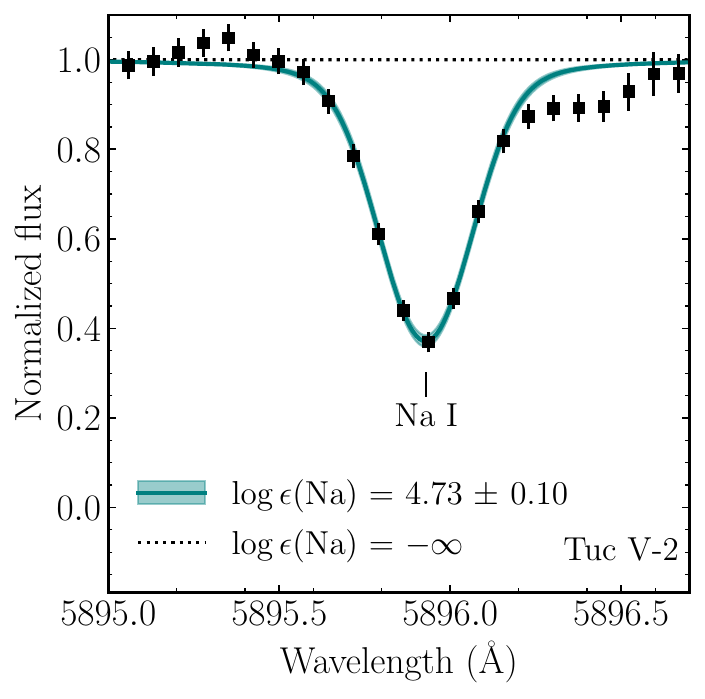}
\includegraphics[width=0.32\linewidth]{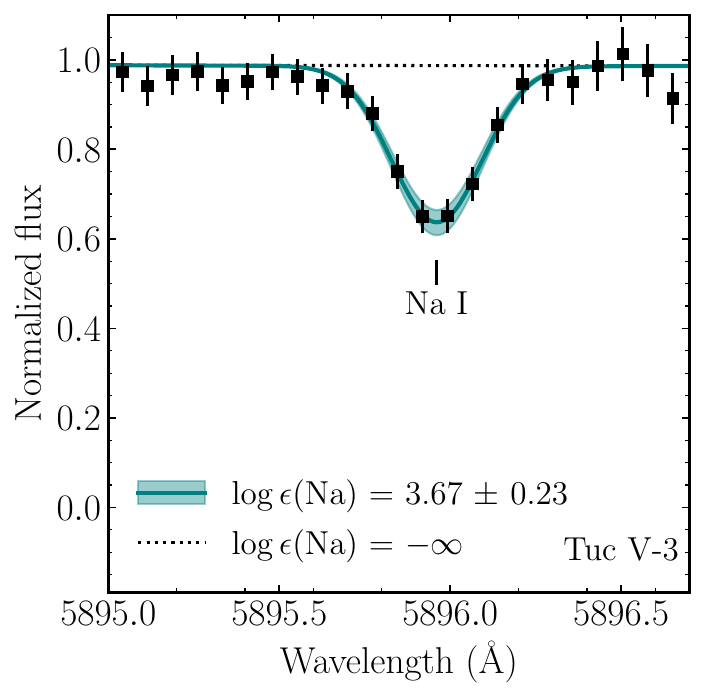}
\caption{\label{fig:synthNa} Comparison of synthesis and observed spectra (black squares) for the 5895\AA\ \ion{Na}{1} line in Tuc~V-1 (left), Tuc~V-2 (middle), and Tuc~V-3 (right). The blue line is the best-fit synthesis, the blue band shows the uncertainty, and the dotted line is a synthesis without the given element.}
\end{figure*}

Carbon abundances were derived via spectral synthesis of the C-H $G$-band at 4313\,{\AA}, as shown in Figure \ref{fig:CH}. As no O abundance could be derived for Tuc~V-3 and Tuc~V-2, we assumed a standard oxygen enhancement for metal-poor stars of $\mathrm{[O/Fe]}=0.4$ when deriving the C abundances from these two stars. We have corrected the C abundances of all three stars for the effects of stellar evolution following \citet{placco2014}. These values are listed in Table \ref{tab:abun} under $C_{corr}$. One of the stars, Tuc~V-1, qualifies as a carbon-enhanced metal-poor (CEMP; \citealt{beers2005}) star, with $\mathrm{[C/Fe]} > 0.7$ \citep{aoki2007}. For this star, we also derive an O abundance from the 6300, 7772, and 7775\,{\AA} lines and an N abundance from the CN band at 3880\,{\AA}. The star is enhanced in both O and N, as has also been seen for MW halo CEMP stars \citep{norris2013}. Na abundances are derived from EW measurements of the 5688\,{\AA}, 5889\,{\AA}, 5895\,{\AA}, 8183\,{\AA}, and 8194\,{\AA} lines and  Al abundances from synthesis of the  3944\,{\AA} and 3961\,{\AA} lines. Figure \ref{fig:synthNa} shows the fit of the 5895\AA, \ion{Na}{1} line in the three stars. The CEMP star, Tuc~V-1, has a very high $\mathrm{[Na/Fe]}$ ratio, a signature which, to some extent, is also present in Tuc~V-2, while Tuc~V-3 is more similar to literature values for UFD galaxy stars (see Figure \ref{fig:abun}). 
K abundances are derived from EW measurements of the 7664\,{\AA} and 7698\,{\AA} lines and Sc abundances from a synthesis of ten lines  (see Table \ref{tab:lines} for details). All three stars have high K abundances ($\mathrm{[K/Fe]} > 0.5$, a unique signature found only in some UFD galaxies \citep{webber2023}.

\subsection{$\alpha$ and iron-peak elements}
\begin{figure*}[hbt!]
\centering
\includegraphics[width=0.32\linewidth]{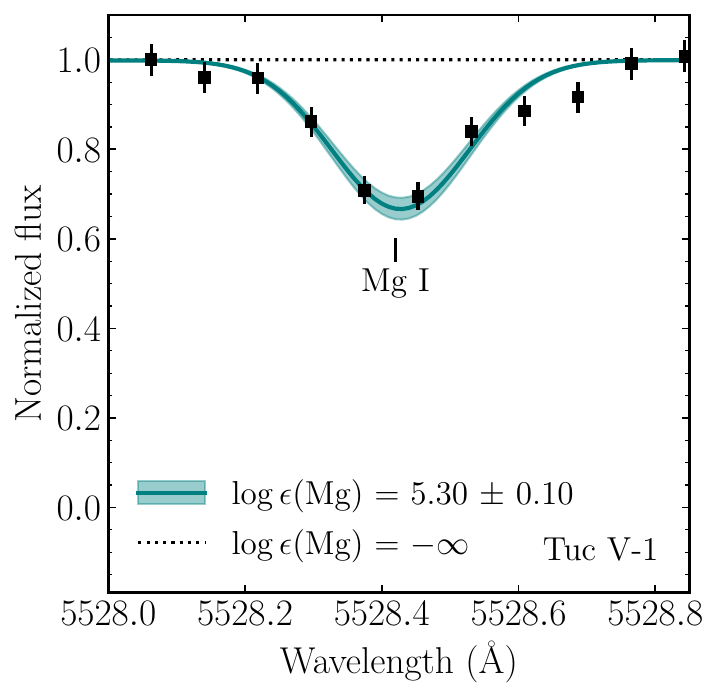}
\includegraphics[width=0.32\linewidth]{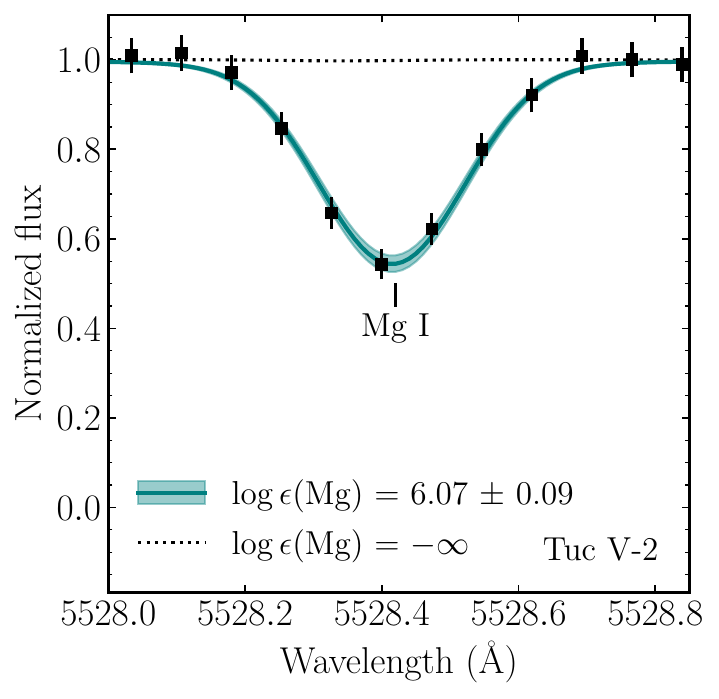}
\includegraphics[width=0.32\linewidth]{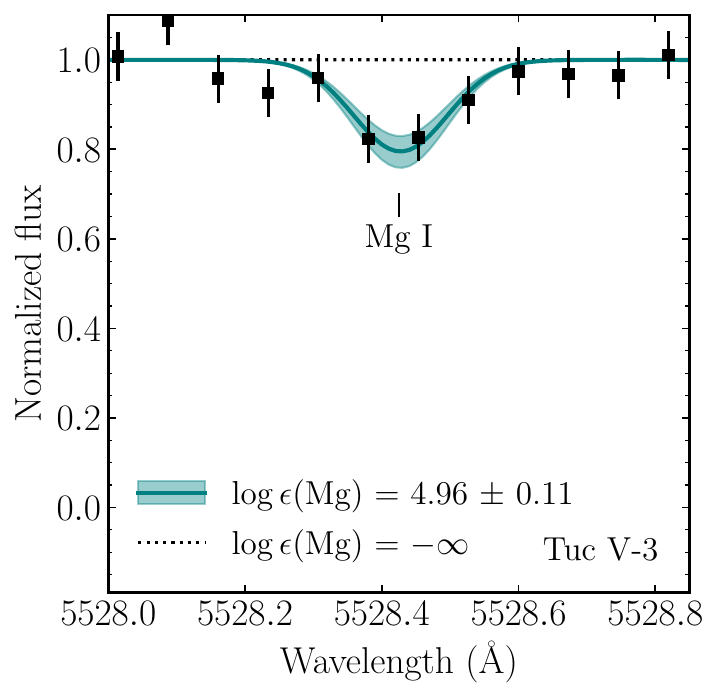}
\caption{\label{fig:synthMg} Comparison of synthesis and observed spectra (black squares) for the 5528\AA\ \ion{Mg}{1} line in Tuc~V-1 (left), Tuc~V-2 (middle), and Tuc~V-3 (right). The blue line is the best-fit synthesis, the blue band shows the uncertainty, and the dotted line is a synthesis without the given element.}
\end{figure*}

Abundances for Mg, Si, Ca, Ti, Cr, and Ni are derived from EW analysis, while abundances of elements with hfs V, Mn, and Co are derived from spectral synthesis (see Table \ref{tab:lines} for details of lines used). Figure \ref{fig:synthMg} shows the fit of the 5528\AA, \ion{Mg}{1} line in the three stars, highlighting the difference in Mg abundance between the stars, especially for Tuc~V-2 and Tuc~V-3. We derive Ti abundances from both neutral and ionized Ti lines and find abundances in agreement within the uncertainties for all three stars, similarly for the \ion{Cr}{1} and \ion{Cr}{2} abundances in Tuc~V-2. As can be seen in Figure \ref{fig:abun}, the $\alpha$ and iron-peak element abundances in Tuc~V mostly follow results from other UFD galaxies, except for Mn, which is very low in all the Tuc~V stars. The Mn abundances in all three stars are derived from the Mn triplet around 4030\AA, which is known to result in lower abundances compared to other Mn lines \citep{cayrel2004,sneden2023}. Unfortunately, no other Mn lines were present in the spectra, supporting the low Mn abundances derived for the stars. A few other notable outliers are the high $\mathrm{[Si/Fe]}$ in the CEMP star, Tuc~V-1, and the high $\mathrm{[Ca/Fe]}$ in Tuc~V-3.

\subsection{Neutron capture elements}
\begin{figure*}[hbt!]
\centering
\includegraphics[width=0.32\linewidth]{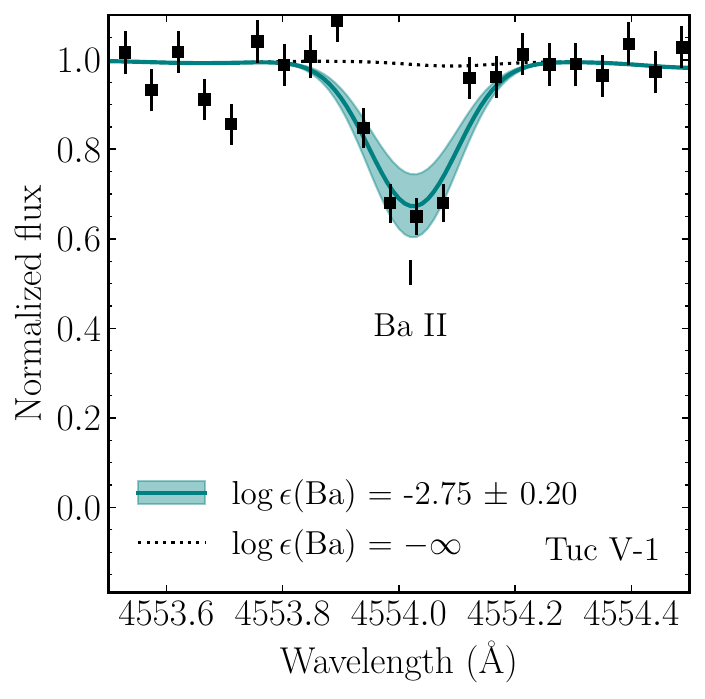}
\includegraphics[width=0.32\linewidth]{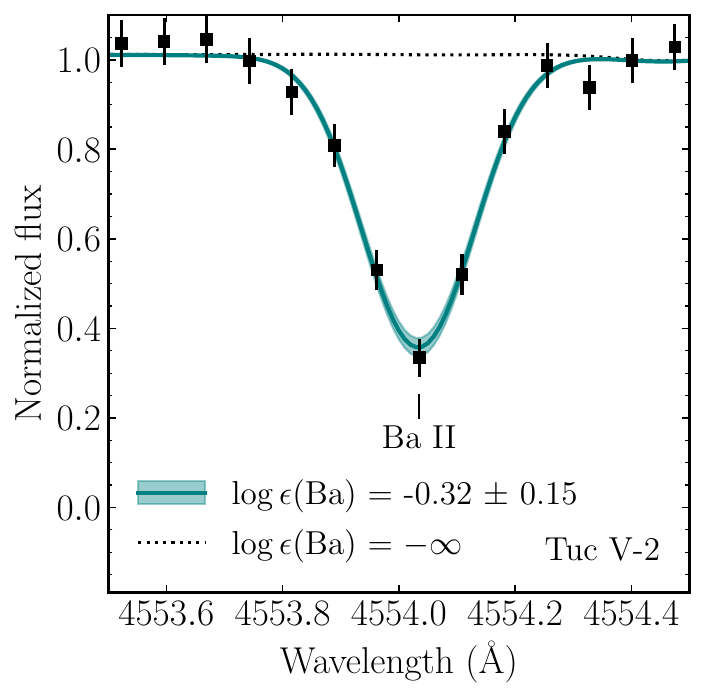}
\includegraphics[width=0.32\linewidth]{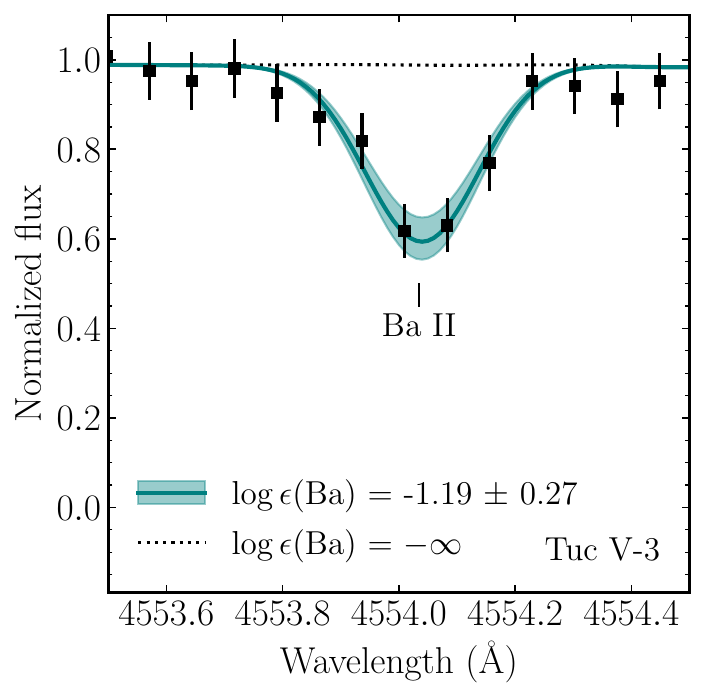}
\caption{\label{fig:synthBa} Comparison of synthesis and observed spectra (black squares) for the 4554\AA\ \ion{Ba}{2} line in Tuc~V-1 (left), Tuc~V-2 (middle), and Tuc~V-3 (right). The blue line is the best-fit synthesis, the blue band shows the uncertainty, and the dotted line is a synthesis without the given element.}
\end{figure*}

All abundances of neutron-capture elements are derived from spectral synthesis. For Tuc~V-1 and Tuc~V-3, we could only derive abundances for Sr and Ba. Both stars exhibit very low $\mathrm{[Sr/Fe]}$ and $\mathrm{[Ba/Fe]}$ ratios characteristic of UFD galaxy stars \citep{ji2019a} and labeling Tuc~V-1 as a CEMP-no star ($\mathrm{[Ba/Fe]} < 0$; \citealt{beers2005}). For Tuc~V-2, however, due to a small enhancement in neutron-capture elements, abundances for Sr, Y, Zr, Ba, La, Pr, Nd, Eu, and Dy could also be derived. The spectral synthesis of the 4554\AA, \ion{Ba}{2} in all three stars is shown in Figure \ref{fig:synthBa} where the higher neutron-capture element abundance of Tuv~2 is clearly visible. Synthesis of additional neutron-capture elements lines in Tuc~V-2 are shown in Figure \ref{fig:synthncap}.

\section{Discussion \label{sec:discus}}

\subsection{Nature of Tuc~V}
\label{sec:nature}

The generally accepted criteria for distinguishing dwarf galaxies from globular clusters include direct dynamical evidence for dark matter from stellar kinematics or indirect evidence for dark matter based on a potential well deep enough to retain supernova ejecta, leading to a substantial spread in iron abundance among member stars \citep{willman2012}.  Although the stellar kinematics of \tucv\ do not place meaningful constraints on its dynamical mass \citep{simon20}, the chemical abundance measurements presented above clearly establish that \tucv\ is a galaxy.  Its stars span a metallicity range from $\mbox{[Fe/H]} = -3.55$ to $\mbox{[Fe/H]} = -2.46$, much larger than has been seen in any faint cluster (and extending to lower metallicities than known clusters).  Other heavy elements such as Ca also vary by more than 1~dex within \tucv, further supporting the system's ability to hold on to supernova ejecta.  The detailed abundances of \tucv\ also resemble those of dwarf galaxies much more than globular clusters: the neutron-capture abundances are very low in two of the three stars ($\mbox{[Sr/Fe]} < -1.5$, $\mbox{[Ba/Fe]} < -1$; \citealt{ji2019a}), the K abundances are high \citep{webber2023}, and CEMP-no stars have not been identified in clusters.  Finally, we note that the metallicities of the Tuc~V stars extend well below any known SMC stars, and the $r$-process abundances are lower than those of the most metal-poor stars in the SMC as well \citep{reggiani21}.  These results argue strongly against the SMC origin suggested by \citet{conn18}, as already concluded by \citet{simon20} based on the velocity of Tuc~V.

With the updated sample of five member stars (four of which have metallicity measurements), we can also redetermine the bulk properties of \tucv.  Using the methodology from \citet{li2017}, we compute a mean metallicity of $\mathrm{[Fe/H]} = -2.84^{+0.32}_{-0.30}$, with a metallicity spread of $\sigma_{\mathrm{[Fe/H]}} = 0.43^{+0.32}_{-0.15}$.  This revised metallicity places \tucv\ close to the stellar mass-metallicity relationship for dwarf galaxies \citep[e.g.,][]{kirby13b}.  Both the metallicity and metallicity spread are in agreement with the photometric measurements of a sample of six \tucv\ stars by \citet{fu2023}.  Using mean velocities for Tuc~V-2 and Tuc~V-3, as well as the center-of-mass velocity from the TheJoker MCMC fit for Tuc~V-1 (\S~\ref{sec:binary}), the systemic velocity for Tuc~V is $-34.7^{+0.9}_{-0.8}$~km~s$^{-1}$, and the velocity dispersion is $\sigma = 1.2^{+0.9}_{-0.6}$~km~s$^{-1}$.  The velocity dispersion is only marginally resolved, and we compute 90\%\ (95.5\%) upper limits of 2.5 (3.1)~km~s$^{-1}$.  These values are generally consistent with, but significantly more precise than, the results of \citet{simon20}; the mean metallicity differs by more than $2\sigma$, but since the original determination relied on only two stars, the uncertainty may have been underestimated.

\subsection{CEMP-no stars in UFD galaxies}
The most metal-poor star in the sample, Tuc~V-1, can be classified as a CEMP-no star \citep{beers2005}, with significant enhancements in C ($\mathrm{[C/Fe]} = 1.77$), N ($\mathrm{[N/Fe]} = 1.81$), and O ($\mathrm{[O/Fe]} = 2.32$), but low abundances of the neutron-capture elements Sr ($\mathrm{[Sr/Fe]} = -1.73$) and Ba ($\mathrm{[Ba/Fe]} = -1.22$). In Figure \ref{fig:CH}, we show the synthesis of the C-H G-band in three sample stars. \citet{ji2020a} compared the fraction of CEMP stars in UFD galaxies at $\mathrm{[Fe/H]} < -2$ to that of the MW halo from \citet{placco2014}. They found that these are essentially identical, around 40\%, while \citet{skuladottir2023} find a much lower fraction of CEMP stars of 9\% below $\mathrm{[Fe/H]} = -3$ in the classical dwarf spheroidal galaxy (dSpH) Sculptor.

Tuc~V-1 also exhibits substantial enhancements in the light elements Na, Mg, and Si, a signature detected in a subset of both the MW halo and UFD CEMP-no stars \citep[e.g.][]{norris2013,spite2018,hayes2023}, and despite the potentially lower fractions of CEMP stars, also in dSph galaxies \citep{hansen2023,roederer2023}. The origin of the peculiar abundance pattern of the CEMP-no stars (both with and without the additional light element enhancement) is still unknown. It has been suggested that this chemical signature can be produced by massive population III (Pop~III) stars ending their lives in low-energy or faint supernovae (SNe) with varying degrees of mixing and fallback \citep[e.g.,][]{umeda2003,iwamoto2005,nomoto2013,ishigaki2014}. Another proposed source is fast-rotating massive metal-free stars, so-called spinstars \citep{meynet2006,liu2021}. Models of spinstars, which include extensive processing and mixing, can reproduce the additional light element enhancement \citep{maeder2015}. More recent theoretical work also suggests that CEMP-no stars can form from gas enriched by Pop~II type II SNe \citep{jeon2021} or asymptotic giant branch (AGB) stars \citep{sharma2019}. In an attempt to characterize the progenitors of CEMP-no stars in UFD galaxies, \citet{rossi2023} constructed a chemical evolution model aiming to match the metallicity distribution function of the Bo\"otes~I galaxy. In their model, they include Pop~III and Pop~II SNe and AGB stars as sources of C and find that CEMP-no stars with $A(\rm C) \leq 6$ were likely formed from gas polluted by Pop~III SNe, whereas Pop~II SNe or Pop~II AGB stars were the primary source of C for stars with $A(\rm C) \geq 6$.

\begin{figure}[hbt!]
\centering
\includegraphics[width=0.99\linewidth]{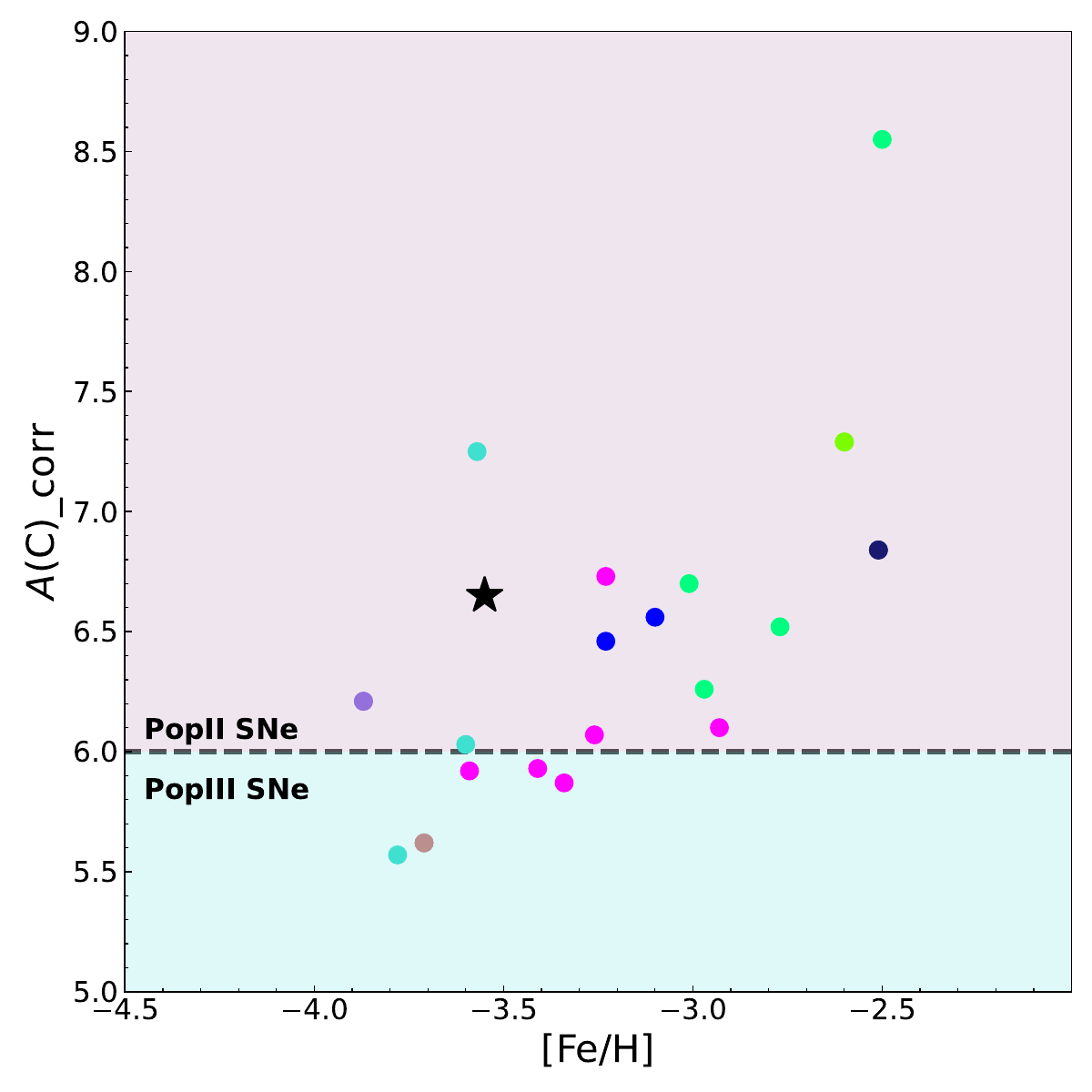}
\includegraphics[width=\linewidth]{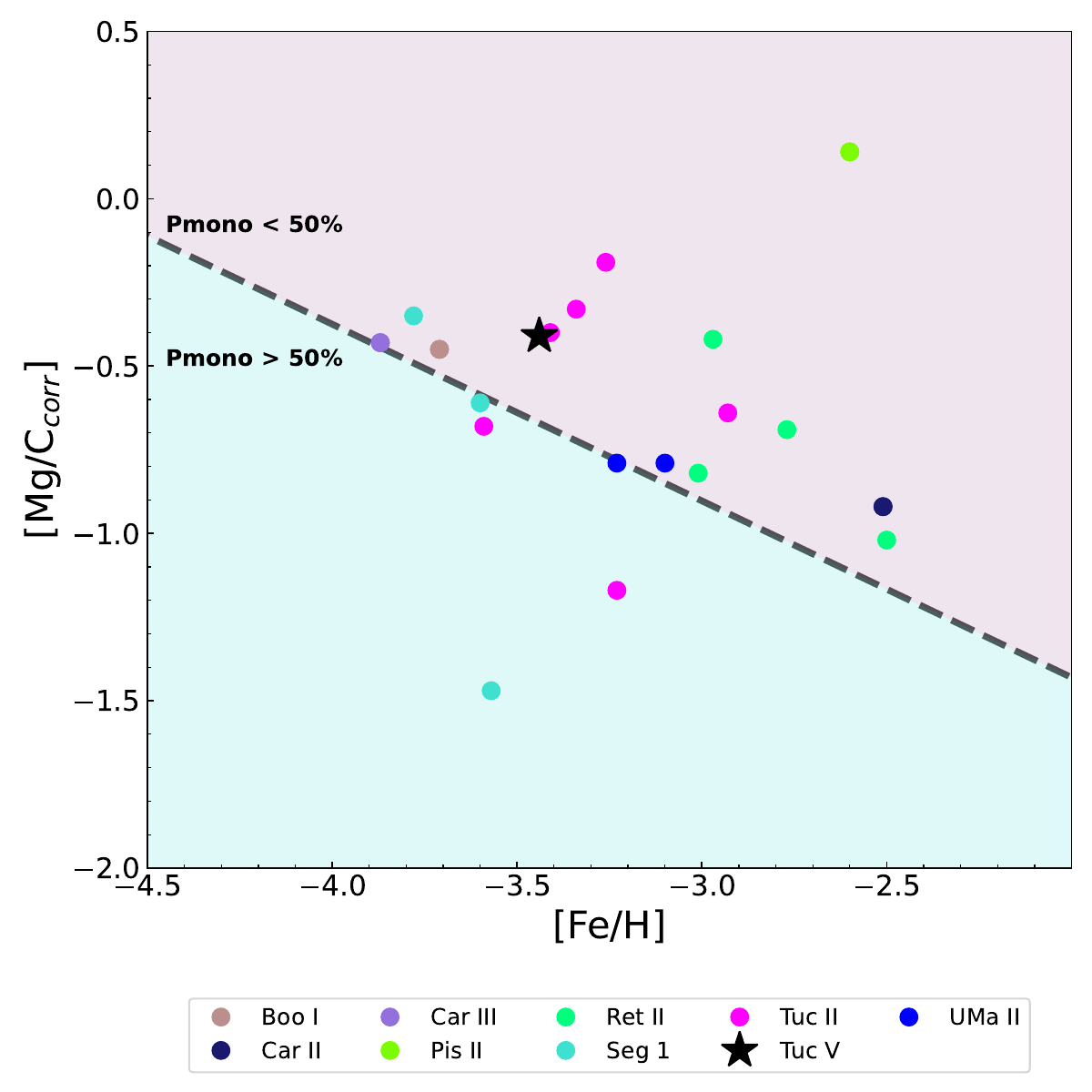}
\caption{Top: Absolute C abundances as a function of metallicity for the CEMP-no stars identified in UFD galaxies. The grey dashed line at $A(\rm C) = 6$ marks the division between enrichment by Pop~II SNe or AGB stars ($A(\rm C) > 6$) and Pop~III SNe ($A(\rm C) < 6$) according to the \citet{rossi2023} model. In this picture, most of the UFD CEMP-no stars, including Tuc~V-1, have $A(\rm C)$ values consistent with enrichment by Pop~II SNe or AGB stars. Bottom: $\mathrm{[Mg/C_{corr}]}$ as a function of metallicity for the UFD galaxy CEMP-no stars. The dashed line separates areas in this abundance space where the probability of stars being mono-enriched is $>50\%$ and $<50\%$ adapted from Figure 11 in \citet{hartwig2018}. According to this metric, the gas from which most of the CEMP-no stars in UFD galaxies, including Tuc~V-1, formed was enriched by multiple progenitors. \label{fig:MgC}} 
\end{figure}

The top panel of Figure \ref{fig:MgC} shows the $A(\rm C)$\footnote{$A(\rm C)$ is corrected for depletion due to stellar evolution.} as a function of metallicity for the CEMP-no stars in UFD galaxies. In this plot, we include four stars from Ret~II, which all have high Ba abundances and thus do not fulfill the $\mathrm{[Ba/Fe]} < 0$ requirement for CEMP-no stars \citep{beers2005}. However, the Ba in these stars is likely of an $r$-process origin \citep{ji2016b,ji2016c,hayes2023} and thus not coupled to the C enhancement. Inspection of this plot shows that Pop II SNe likely enriched most CEMP-no stars in UFD galaxies (purple shaded region), according to the model from \citet{rossi2023}. In principle, Pop~II AGB stars could also have polluted the gas from which these stars formed. However, since all of these stars have very low Ba abundances, $\mathrm{[Ba/Fe]} < -0.5$\footnote{Except for the Ret~II stars as explained above.}, these AGB stars would need to have a very low or non-existent production of $s$-process elements, like Ba during the AGB phase.

Another test of the origin of CEMP-no stars can be made using the results from \citet{hartwig2018}, who argued that the $\mathrm{[Mg/C]}$ ratio of stars could be used to determine if metal-poor stars have been enriched by single (mono-enriched) or multiple (multi-enriched) progenitors, where stars with $\mathrm{[Mg/C]} < -1$ are most likely to be mono-enriched. In the bottom panel of Figure \ref{fig:MgC}, we plot the $\mathrm{[Mg/C_{corr}]}$ as a function of metallicity for the UFD galaxy CEMP-no stars, and a dashed line separate the areas in this abundance space where the probability of stars being mono-enriched is $>50\%$ (cyan shaded region) and $<50\%$ (purple shaded region) adapted from Figure 11 in \citet{hartwig2018}. As can be seen, the models from \citet{hartwig2018} and \citet{rossi2023} largely agree on the fraction of stars enriched by Pop~II SNe (multi-enriched) and by Pop~III SNe (mono-enriched). However, the two methods somewhat disagree on which exact stars are likely mono-enriched. The highly C-enhanced star in Seg~1 highlights the divergence of the models, suggesting that using just C as a measure for enrichment might not be sufficient. It should be mentioned, though, that neither \citet{hartwig2018} nor \citet{rossi2023} include spinstars as a source of C and Mg in their models. Hence, these objects still cannot be ruled out as possible progenitors of CEMP-no stars.

\subsubsection{Binarity of Tuc~V-1}
\label{sec:binary}
As mentioned in Section \ref{sec:obs} Tuc~V-1 is part of a binary system. To determine the orbital parameters of this system, we fit the velocity data listed in Table~\ref{tab:obs} using the rejection sampling algorithm TheJoker \citep{pricewhelan17}.  TheJoker returned a likely period of $\sim380$~d. In order to explore the likelihood around this solution more completely, we then carried out a Markov Chain Monte Carlo (MCMC) fit to the orbital parameters using PyMC.  Based on the results from TheJoker, we initialized the MCMC with a prior on the period that was flat between 320~d and 405~d, and we set Gaussian priors on the velocity semi-amplitude and center-of-mass velocity using the best-fit TheJoker values and increasing the standard deviation on each quantity by a factor of $\sim5$ so as not to unduly restrict the parameter space.  The posterior parameter distributions from the MCMC were: $P = 381^{+5}_{-4}$~d, $e = 0.10^{+0.03}_{-0.10}$, $K = 11.0^{+0.8}_{-1.0}$~km~s$^{-1}$, and $v_{hel} = -32.4^{+0.7}_{-0.8}$~km~s$^{-1}$.  The resulting binary mass function is $f = 0.05 \pm 0.01$~M$_{\odot}$, corresponding to a minimum companion mass of 0.4~M$_{\odot}$.  Depending on the unknown inclination angle of the system, this measurement is consistent with an M dwarf, K dwarf, or white dwarf companion star.

\begin{figure}[hbt!]
\centering
\includegraphics[width=\linewidth]{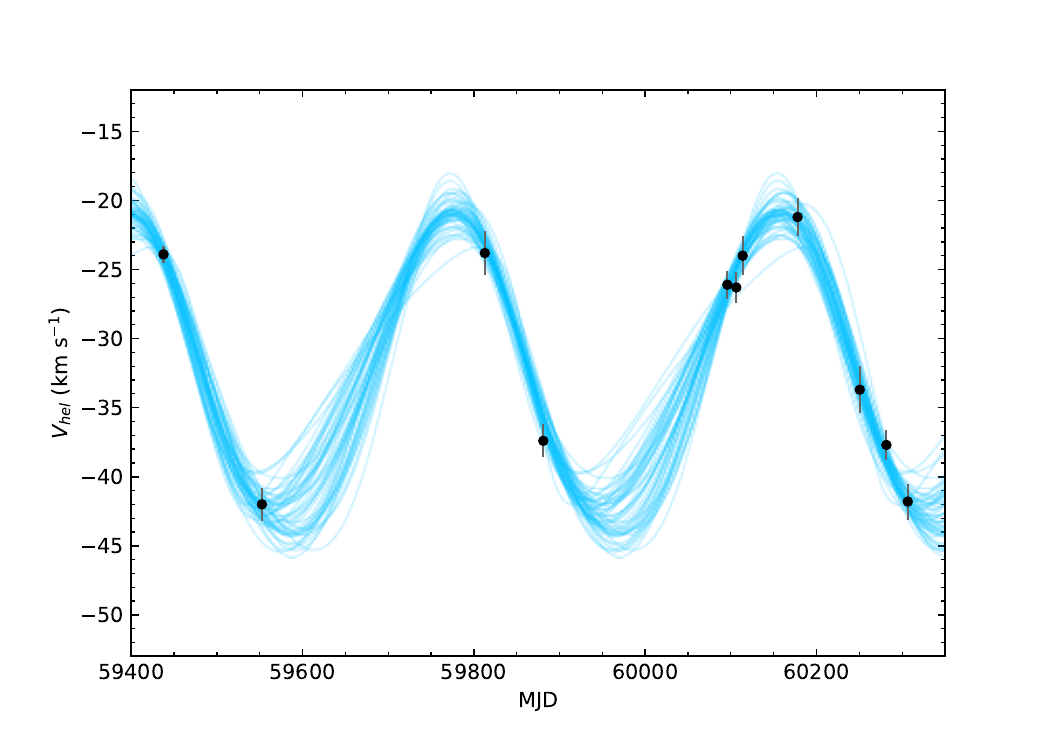}
\caption{\label{fig:orbit} Radial velocity measurements for Tuc~V-1 over the $\sim2$~years since our initial observation.  The blue lines show a randomly selected set of MCMC samples that fit the velocity data.}
\end{figure}

The fact that Tuc~V-1 is part of a binary system introduces the possibility that mass transfer from a more massive evolved companion has altered the abundances of this star, being the source of the C enrichment. However, extensive radial-velocity monitoring of CEMP-no stars in the halo has found the majority of these to be single \citep{starkenburg2014,hansen2016}. Thus, these stars likely inherited their chemical peculiarities from the cloud they formed from rather than through binary interaction. Still, there is some observational evidence suggesting that the binary fraction of CEMP-no stars in the halo and dSph galaxies is correlated with the C abundance, with stars in binary systems having higher absolute C abundances than single CEMP-no stars \citep{arentsen2019,hansen2023}.

One other UFD CEMP-no star, J033607 in Ret~II, has been found to show radial-velocity variations \citep{ji2023,hayes2023}. Of the remaining UFD galaxy CEMP-no stars, 12 do not show radial velocity variations \citep{frebel2010,norris2010,spite2018,chiti2023,ji2023}, while seven do not have multiple velocities reported in the literature. Looking at the absolute C abundances, Tuc~V-1 has $A(C)= 6.65$, and the Ret~II binary star has $A(C)= 6.21$, which, when inspecting the top panel of Figure \ref{fig:MgC}, places these two stars in the middle or lower end of the $A(\rm C)$ distribution for the UFD galaxy CEMP-no stars, thus not following the trend seen for the halo and dSph where binary CEMP-no stars are found mostly at $A(C) > 7$ \citep{arentsen2019,hansen2023}.

\subsection{The $r$-process in UFD galaxies}
\begin{figure}[hbt!]
\centering
\includegraphics[width=\linewidth]{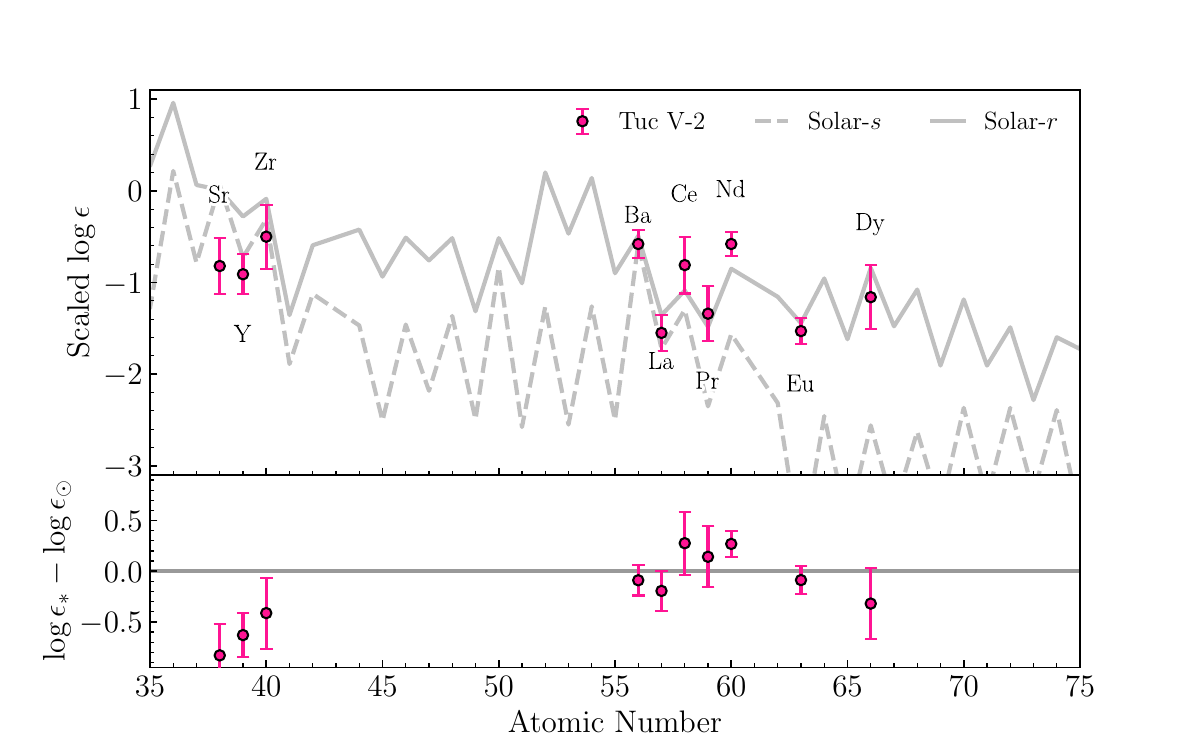}
\caption{\label{fig:rpro} Absolute abundances for neutron-capture elements derived for Tuc~V-2 compared to scaled Solar system $r-$ (solid) and $s$-process (dashed) abundances taken from \citep{sneden2008}. The stellar abundances most closely match the Solar system $r$-process abundance pattern.}
\end{figure}

The most metal-rich star in our sample, Tuc~V-2, exhibits a small enhancement in some neutron-capture elements. It has $\mathrm{[Eu/Fe]} = 0.36$ and $\mathrm{[Ba/Fe]} = -0.32$, resulting in a Ba to Eu ratio of -0.68, consistent with pollution by a rapid neutron-capture ($r-$)process \citep{sneden2008}. In Figure \ref{fig:rpro}, we plot the absolute neutron-capture element abundances derived for the star and compare them to the Solar-System $r-$ and $s$-process abundance patterns scaled to the stellar Ba and Eu abundances, respectively. Looking at the heavy neutron-capture elements (Ba and up), it is clear from this figure that the $r$-process pattern best matches the data of the two, confirming the $r$-process origin of the neutron-capture elements detected in Tuc~V-2. For the lighter elements  Sr, Y, and Zr, the agreement with the Solar pattern is less good; a similar signature has also been seen among MW halo $r$-process enhanced stars and has been suggested to be the result of a limited $r$-process operating in this element range \citep{mcwilliam1998,hansen2012,roederer2022}.

$r$-process-enhanced stars have been found in three other UFD galaxies, with 72\% of the stars analyzed in Reticulum~II being highly $r$-process enhanced ($r$-II stars) \citep{ji2023}, while the five stars analyzed in Tucana~III and one star in Grus~II are mildly enhanced (0.22 < $\mathrm{[Eu/Fe]}$ < 0.60), and labeled $r$-I stars \citep{hansen2017,marshall2019,hansen2020}. 

The astrophysical site(s) of the $r$-process is still debated, but currently only neutron star mergers (NSMs) have been observationally confirmed as an $r$-process element production site with the GW170817 event \citep{chornock2017,drout2017,pian2017,tanvir2017}, although other sites, including various exotic SNe, have been proposed \citep{winteler2012,mosta2018,grichener2019,siegel2019a}. However, the lanthanide fraction of the material produced in GW170817 has been shown not to agree with the lanthanide fraction of the most $r$-process enriched MW halo and UFD galaxy stars, with the ejecta from GW170817 not being lanthanide rich enough to match the composition found in the stars \citep{ji2019b}. 

Currently, different NSM models predict different ejecta compositions. For example, models from \citet{eichler2015} and \cite{lippuner2017} find the dynamical ejecta to be neutron-rich, producing a high lanthanide fraction, whereas wind ejecta, for example, from an accretion disk, contributes more of the light $r$-process elements (below the second peak), while in models from \citet{radice2018}, $r$-process material with $90 < A < 125$ is also produced in the dynamical ejecta, depending on the total mass of the neutron stars and the equation of state. For GW170817, most of the ejected material came from the post-merger accretion disk \citep[and references therein]{siegel2019b}, suggesting that mergers with dynamical ejecta constituting a larger fraction of the total ejecta mass might be needed to match the stellar abundances.

A simple way to investigate the ratio of lanthanide-poor to lanthanide-rich material is by looking at the $\mathrm{[Sr/Eu]}$ ratios of the stars. For Tuc~V-2, we measure $\mathrm{[Sr/Eu]} = - 1.62$, very similar to the value for the $r$-I star in Gru~II ($\mathrm{[Sr/Eu]}= -1.44$) and the average of the seven $r$-II stars in Ret~II ($\mathrm{[Sr/Eu]}= -1.44$), while the average value for the five $r$-I stars in Tuc~III  is somewhat higher at $\mathrm{[Sr/Eu]}= -0.58$. The fairly consistent values found for Gru~II, Ret~II, and Tuc~V could suggest that the (potential) NSMs polluting these galaxies shared some characteristics.  

\citet{holmbeck2021} used the $\mathrm{[Zr/Dy]}$ ratios of $r$-process-enhanced MW halo stars to backward-model the properties of the NSMs required to create the measured abundance ratios. They found that asymmetric systems ($M_1 \ll M_2$), which have higher dynamical ejecta mass, were generally needed to reproduce the lowest $\mathrm{[Zr/Dy]}$ ratios, which are equivalent to the low $\mathrm{[Sr/Eu]}$ derived for the Gru~II, Ret~II, and Tuc~V $r$-process-enhanced stars. 

The time delay connected to the NSM scenario, however, still introduces a problem with this enrichment channel. For example, for Ret~II the large fraction of $r$-process enriched stars combined with the star formation history of the galaxy \citep{simon2023}, led \citet{ji2023} to suggest that a rare type of core-collapse SNe is more likely the source of the $r$-process material in this galaxy than a NSM. In both Gru~II and Tuc~V, it is the most metal-rich star in the sample that exhibits the $r$-process enrichment, suggesting that the $r$-process production occurred after a delay, late in the star formation process in these galaxies. The analysis of future larger samples of stars in these two systems providing information on the overall $r$-process enhancement level of the galaxies and their star formation histories, like in Ret~II, will reveal more about their $r$-process element polluters.

\subsection{Chemical diversity in UFD galaxies}
The three Tuc~V stars analyzed in this paper present very diverse sets of chemical compositions. In addition to the carbon and $r$-process enrichment discussed above, the three stars in Tuc~V span a remarkably wide range in their Mg to Ca ratios with $\mathrm{[Mg/Ca]} = 0.89$ (Tuc~V-1), $\mathrm{[Mg/Ca]} = 0.29$ (Tuc~V-2), and $\mathrm{[Mg/Ca]} = -0.75$ (Tuc~V-3).  Excluding the CEMP-no star (Tuc~V-1) where the high $\mathrm{[Mg/Fe]}$ value is likely coupled to its C enhancement, the $\mathrm{[Mg/Ca]}$ ratios for the other two stars still differ by 1.04~dex despite their similar metallicities, which suggest a decoupling between the Fe and $\alpha$ enrichment of the stars. Furthermore, such large variations in enrichment, especially among stars with similar metallicities, imply that the interstellar medium of \tucv\ must have been highly inhomogeneous and, correspondingly, that the number of SNe responsible for the chemical enrichment must have been small. 

Ranges in $\mathrm{[Mg/Ca]}$ ratios in UFD galaxy stars are usually interpreted as a consequence of core-collapse SNe of varying masses enriching in the galaxy \citep{hansen2020,ji2020b}. However, Type Ia SNe are also known to produce significant amounts of Ca \citep{kobayashi2020}. Hence, a low $\mathrm{[Mg/Ca]}$ ratio can also be the signature of enrichment by a Type Ia SN. 

Disregarding CEMP-no stars with large enhancements in light elements, the only other UFD galaxy displaying a similar range in $\mathrm{[Mg/Ca]}$ values is Tuc~II, with a range of 1.37~dex, driven by two of the new members discovered in the outskirts of the galaxy and analyzed by \citet{chiti2023}. \citet{chiti2023} argues that Tuc~II was mainly enriched by low energy core-collapse SNe owing to the large fraction of CEMP stars in the galaxy. Still, they also find evidence for enrichment by a sub-Chandrasekhar-mass Type Ia SN in the most metal-rich star in their sample. This star is also the one with the lowest $\mathrm{[Mg/Ca]}$ ratio of -0.72 in their sample, and they argue that some of the Ca in this star likely comes from the Type Ia SN, as Ca does not follow the trend of the other $\alpha$-elements in this star. 

The low $\mathrm{[Mg/Ca]}$ ratio found for Tuc~V-3 in this analysis is mainly driven by a high $\mathrm{[Ca/Fe]}$. This could suggest that Tuc~V-3 has been enriched by a Type Ia SN and possibly also a sub-Chandrasekhar-mass Type Ia SN, as no significant difference is seen between this star and the other two stars for other elements expected to show signatures of Type Ia enrichment like Cr, Mn, and Ni \citep{kobayashi2020}. 

Finally, it has also been suggested that the $\mathrm{[Mg/Ca]}$ ratios of stars in UFD galaxies are coupled to the environment of the galaxy where UFD galaxies associated with the Large Magellanic Clouds (LMC) display a correlation between $\mathrm{[Mg/Ca]}$ and metallicity. With the sample of stars analyzed in this work, it is not possible to determine if a trend is present for $\mathrm{[Mg/Ca]}$ in Tuc~V. However, the orbit of Tuc~V does not indicate it is a satellite of the LMC \citep{simon20}.

\section{Summary} \label{sec:summary}
We present a detailed chemical analysis of three stars in the Tuc~V system. In addition to a bright, known member in the center of the galaxy, Tuc~V-3, we have identified two new members, Tuc~V-1 and Tuc~V-2, in the outskirts of the galaxy from Gaia astrometry. We then use the abundances to investigate the nature and chemical enrichment of the system. The large scatter between the stars detected for abundances of elements like Mg, Ca, Sc, and Fe tells us that Tuc~V is a UFD galaxy with a very inhomogeneous chemical enrichment history. It holds both an $r$-I star (Tuc~V-2) and a CEMP-no star (Tuc~V-1). The C and Mg abundances of the CEMP-no star combined with C and Mg abundances from all known CEMP-no stars in UFD galaxies suggest that the majority of these have experienced enrichment from multiple progenitors. The $\mathrm{[Sr/Eu]}$ ratio of the $r$-I star points to an $r$-process element source that shares characteristics with the sources enriching Gru~II and Ret~II. Finally, we detect a very wide range in the $\mathrm{[Mg/Ca]}$ ratios of the stars. A similar range is only seen in one other UFD galaxy, namely Tuc~II, where the Mg and Ca abundances of stars located some distance from the center drive this diversity. Hence, obtaining the abundances of stars both in the center and in the outskirts of these galaxies is needed to fully assess their chemical diversity.

\acknowledgments
The authors thank the referee, Dr. Ricardo Schiavon for the careful read of the paper and useful comments, which have improved the manuscript. T.T.H acknowledges support from the Swedish Research Council (VR 2021-05556). 
A.P.J. acknowledges support by the National Science Foundation under grants AST-2206264 and AST-2307599.
T.S.L. acknowledges financial support from Natural Sciences and Engineering Research Council of Canada (NSERC) through grant RGPIN-2022-04794.
J.Y.G. acknowledges support from their Carnegie Fellowship.
This paper includes data gathered with the 6.5 meter Magellan Telescopes located at Las Campanas Observatory.
This research made extensive use of the SIMBAD database operated at CDS, Straasburg, France \citep{wenger2000}, \href{https://arxiv.org/}{arXiv.org}, and NASA's Astrophysics Data System for bibliographic information.  We thank Adrian Price-Whelan for assistance with binary fitting using TheJoker.

This work has made use of data from the European Space Agency (ESA) mission {\it Gaia} (\url{https://www.cosmos.esa.int/gaia}), processed by the {\it Gaia} Data Processing and Analysis Consortium (DPAC, \url{https://www.cosmos.esa.int/web/gaia/dpac/consortium}). Funding for the DPAC has been provided by national institutions, in particular, the institutions participating in the {\it Gaia} Multilateral Agreement.

This project used public archival data from the Dark Energy Survey (DES). Funding for the DES Projects has been provided by the U.S. Department of Energy, the U.S. National Science Foundation, the Ministry of Science and Education of Spain, the Science and Technology Facilities Council of the United Kingdom, the Higher Education Funding Council for England, the National Center for Supercomputing Applications at the University of Illinois at Urbana-Champaign, the Kavli Institute of Cosmological Physics at the University of Chicago, the Center for Cosmology and Astro-Particle Physics at the Ohio State University, the Mitchell Institute for Fundamental Physics and Astronomy at Texas A\&M University, Financiadora de Estudos e Projetos, Funda{\c c}{\~a}o Carlos Chagas Filho de Amparo {\`a} Pesquisa do Estado do Rio de Janeiro, Conselho Nacional de Desenvolvimento Cient{\'i}fico e Tecnol{\'o}gico and the Minist{\'e}rio da Ci{\^e}ncia, Tecnologia e Inova{\c c}{\~a}o, the Deutsche Forschungsgemeinschaft, and the Collaborating Institutions in the Dark Energy Survey.  The Collaborating Institutions are Argonne National Laboratory, the University of California at Santa Cruz, the University of Cambridge, Centro de Investigaciones Energ{\'e}ticas, Medioambientales y Tecnol{\'o}gicas-Madrid, the University of Chicago, University College London, the DES-Brazil Consortium, the University of Edinburgh, the Eidgen{\"o}ssische Technische Hochschule (ETH) Z{\"u}rich, Fermi National Accelerator Laboratory, the University of Illinois at Urbana-Champaign, the Institut de Ci{\`e}ncies de l'Espai (IEEC/CSIC), the Institut de F{\'i}sica d'Altes Energies, Lawrence Berkeley National Laboratory, the Ludwig-Maximilians Universit{\"a}t M{\"u}nchen and the associated Excellence Cluster Universe, the University of Michigan, the National Optical Astronomy Observatory, the University of Nottingham, The Ohio State University, the OzDES Membership Consortium, the University of Pennsylvania, the University of Portsmouth, SLAC National Accelerator Laboratory, Stanford University, the University of Sussex, and Texas A\&M University. Based in part on observations at Cerro Tololo Inter-American Observatory, National Optical Astronomy Observatory, which is operated by the Association of Universities for Research in Astronomy (AURA) under a cooperative agreement with the National Science Foundation.

\facility{Magellan:Clay, Magellan:Baade}
\software{MOOG \citep{sneden1973,sobeck2011}, IRAF \citep{tody1986,tody1993}, ATLAS9 \citep{castelli2003}, linemake \citep{placco2021}, NumPy \citep{numpy}, Matplotlib \citep{matplotlib}, AstroPy \citep{Astropy:13,Astropy:18}, CarPy \citep{kelson2003}, 
SMHR \citep{casey2014}}

\appendix
\counterwithin{figure}{section}
\counterwithin{table}{section}

\section{Stellar parameters}
Temperatures used in the work are derived photometrically using the color-$T_{{\rm eff}}$ relations from \citet{casagrande2010}. In Table \ref{tab:teff}, we list temperatures derived from the individual color bands used along with spectroscopic temperatures of the stars corrected for the offset between spectroscopic and photometric temperature scales using the method outlined in \citet{frebel2013}.  The uncertainty related to this correction is 150~K. Table \ref{tab:paramerr} lists the stellar parameter uncertainties arising from the standard deviation of $T_{\rm eff}$ from the four colors used (photo), and the effect of the standard deviation of abundances of \ion{Fe}{1} lines used (stat) as described in section \ref{sec:analysis} 

\begin{deluxetable}{lccccc}[htb]
\caption{Photometric Temperatures\label{tab:teff}}
\tablehead{ID & $T_{\rm eff}(B-V)$ & $T_{\rm eff}(V-R)$ & $T_{\rm eff}(R-I)$ & $T_{\rm eff}(V-I)$ & $T_{\rm eff}(spec)$\\
&(K)&(K)&(K)& (K) & (K) }
\startdata
Tuc~V-1 & 4287 & 4365 & 4512 & 4356 & 4481 \\ 
Tuc~V-2 & 4691 & 4766 & 4861 & 4776 & 4827 \\ 
Tuc~V-3 & 4961 & 4985 & 5015 & 4967 & 5164 \\ 
\enddata
\end{deluxetable}

\begin{deluxetable}{lccc}[htb]
\caption{Stellar parameter uncertainties\label{tab:paramerr}}
\tablehead{Param & Tuc~V-1 & Tuc~V-2 & Tuc~V-3  }
\startdata
$\Delta T_{\rm eff, photo}$ & 94   & 70   & 24  \\ 
$\Delta T_{\rm eff, stat}$  & 45   & 60   & 128  \\ 
$\Delta \log g_{\rm photo}$ & 0.29 & 0.19 & 0.10 \\
$\Delta \log g_{\rm stat}$  & 0.14 & 0.10 & 0.17 \\
$\Delta \xi_{\rm photo}$    & 0.02 & 0.01 & 0.00 \\
$\Delta \xi_{\rm stat}$     & 0.09 & 0.07 & 0.13 \\
$\Delta \mathrm{[Fe/H]}_{\rm photo}$ & 0.14 & 0.03 & 0.03 \\
$\Delta \mathrm{[Fe/H]}_{\rm stat}$ & 0.15 & 0.15 & 0.25 \\
\enddata
\end{deluxetable}

\begin{figure*}[hbt!]
\centering
\includegraphics[width=\linewidth]{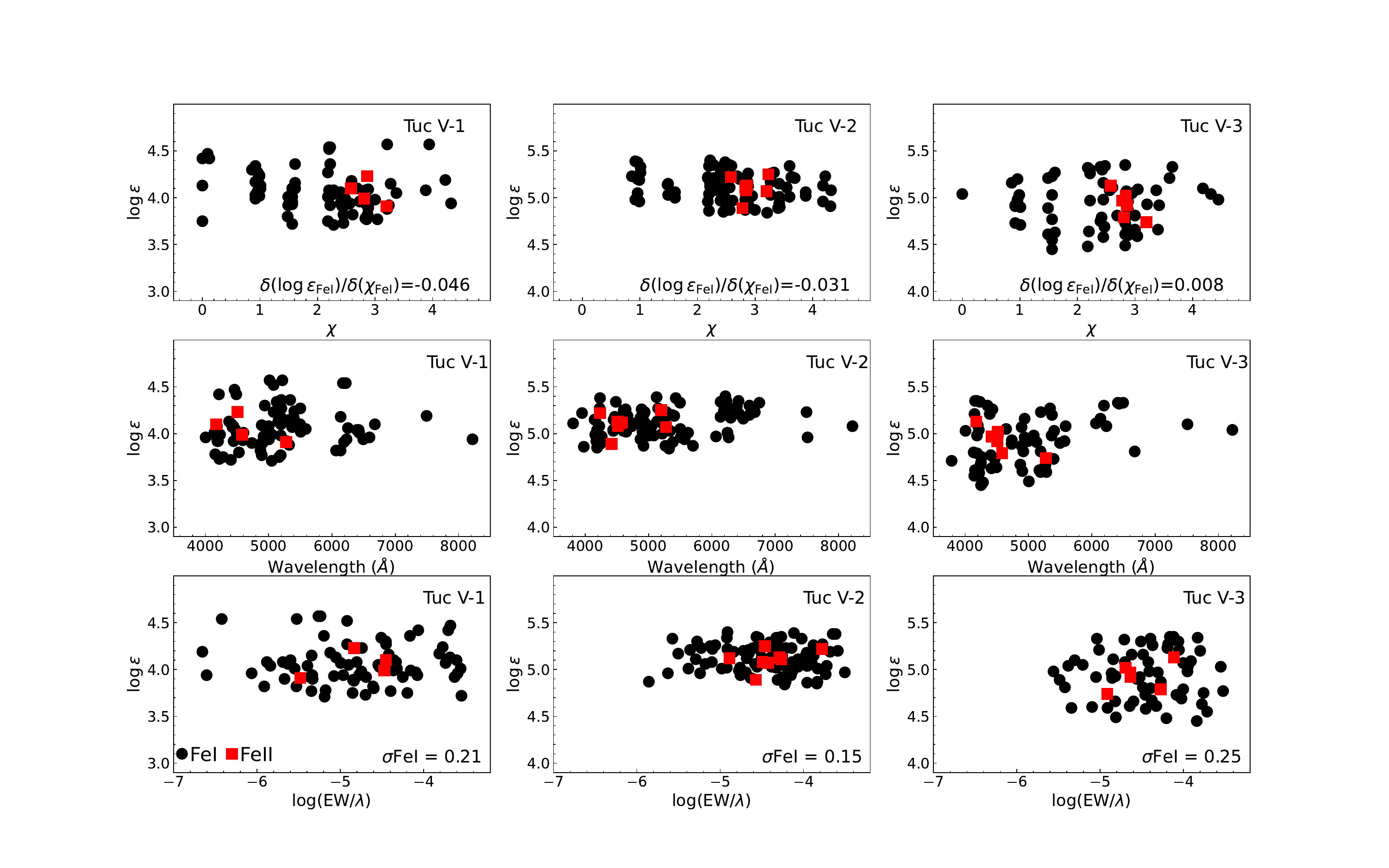}
\caption{\label{fig:param} $\log\epsilon$ of \ion{Fe}{1} and \ion{Fe}{2} lines used for parameter determination of Tuc~V-1 (left), Tuc~V-2 (middle),  and Tuc~V-3 (right) as a function of excitation potential (top), wavelength (middle), and reduced EW (bottom)}
\end{figure*}

\section{Synthesis of neutron-capture elements}

\begin{figure*}[hbt!]
\centering
\includegraphics[width=0.33\linewidth]{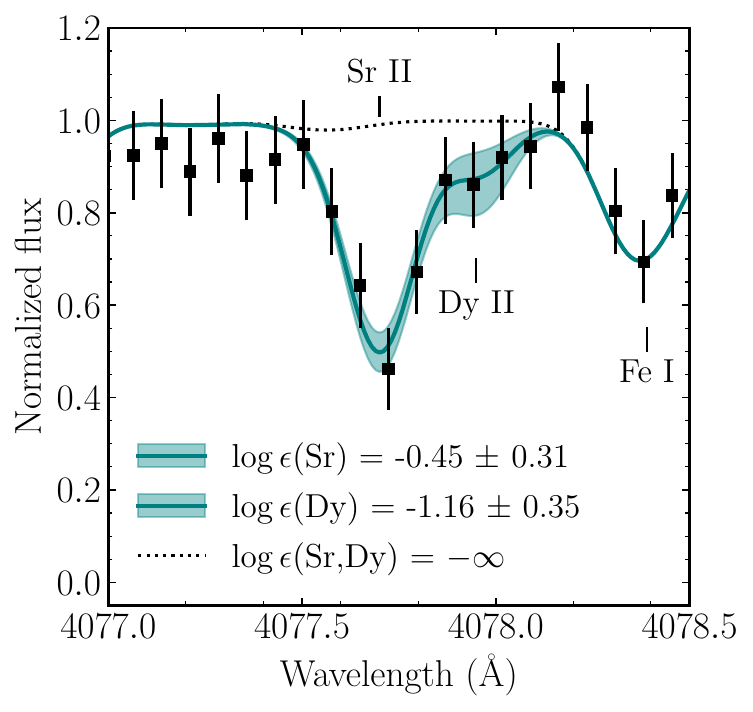}
\includegraphics[width=0.31\linewidth]{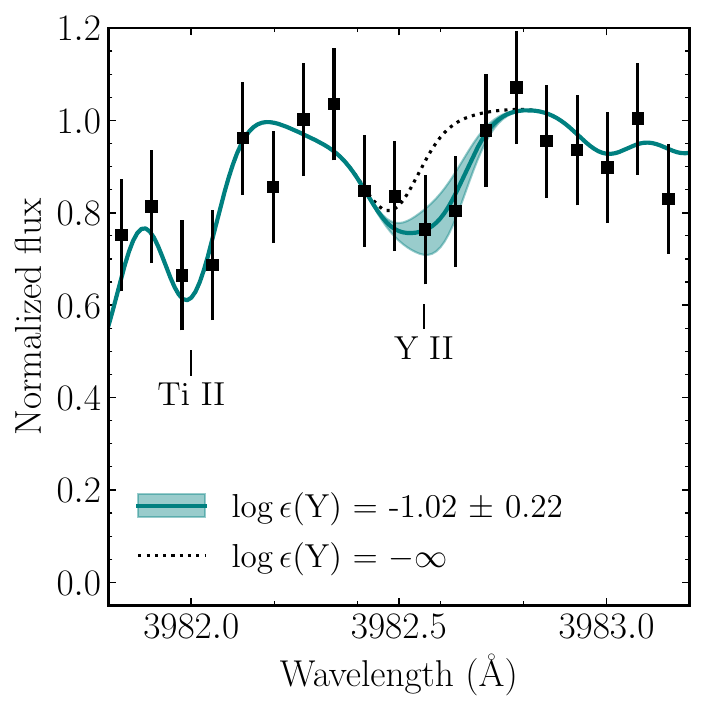}
\includegraphics[width=0.33\linewidth]{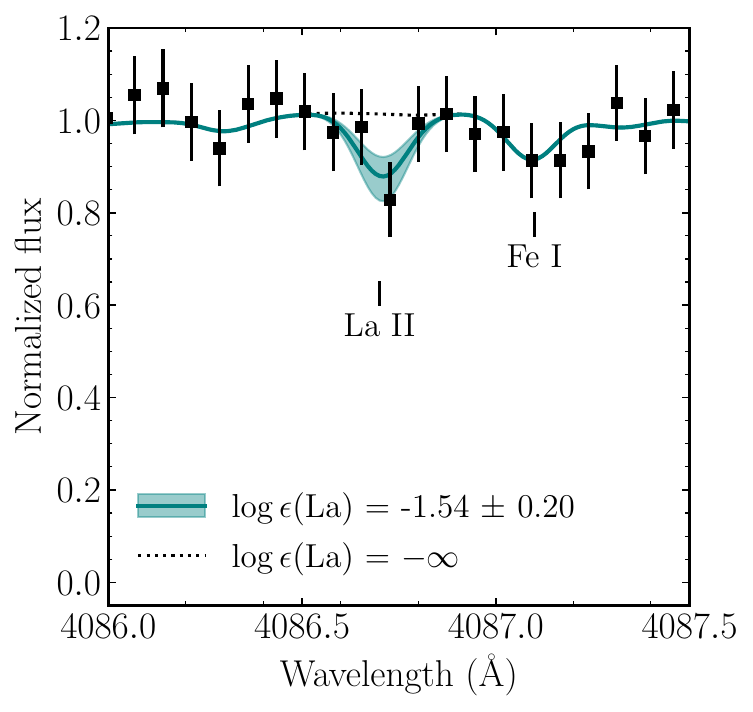}\\
\includegraphics[width=0.31\linewidth]{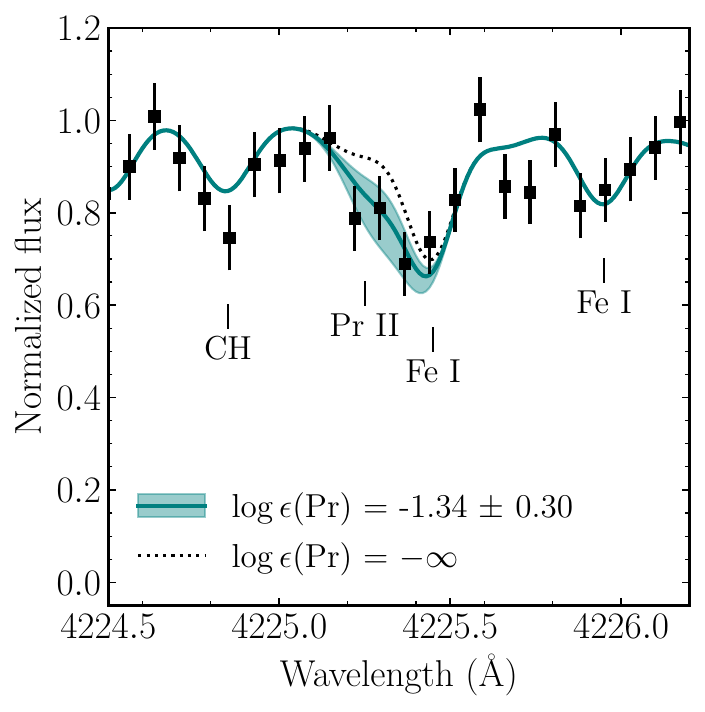}
\includegraphics[width=0.31\linewidth]{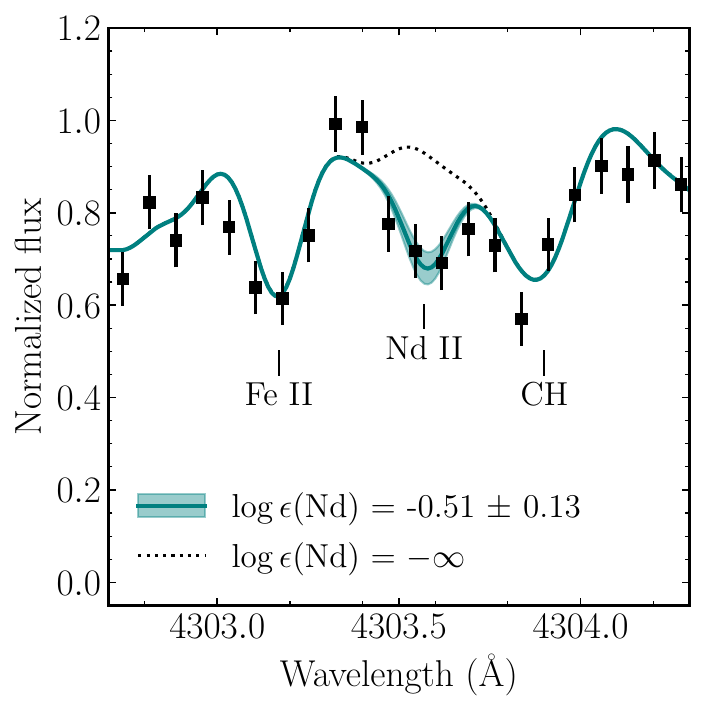}
\includegraphics[width=0.33\linewidth]{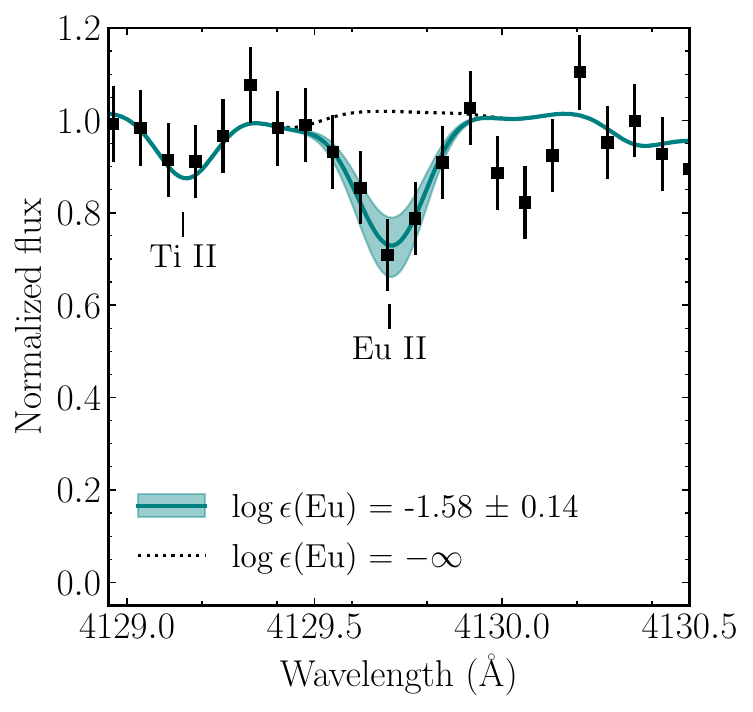}
\caption{\label{fig:synthncap} Comparison of synthesis and observed spectra (black squares) for neutron-capture elements in Tuc~V-2. The blue line is the best-fit synthesis, the blue band shows the uncertainty, and the dotted line is a synthesis without the given element.}
\end{figure*}


\end{document}